\documentclass[aps,prd,reprint,onecolumn,notitlepage,superscriptaddress]{revtex4-1}

\usepackage{lipsum}
\usepackage{amsmath}
\usepackage{graphicx}
\usepackage{color}
\usepackage{tikz}
\usepackage{verbatim}
\usepackage{hyperref}
\usepackage{amssymb}
\usepackage{bm}
\newcommand{\rom}[1]{\textup{\uppercase\expandafter{\romannumeral#1}}}

\begin{document}

\title{Hamiltonian Analysis of Nonlocal F(R) Gravity Models}
\author{Pawan Joshi\footnote{email:pawanjoshi697@iiserb.ac.in}
       }

\affiliation{Indian Institute of Science Education and Research Bhopal,\\ Bhopal 462066, India}

\author{Sukanta Panda \footnote{email:sukanta@iiserb.ac.in}
       }

\affiliation{Indian Institute of Science Education and Research Bhopal,\\ Bhopal 462066, India}

\begin{abstract}
We construct a Hamiltonian for the nonlocal F(R) theory in the present work. By this construction, we demonstrate the nature of the ghost degrees of freedom. Finally, we find conditions that give rise to ghost-free theories. 

\end{abstract}

\maketitle

\section{Introduction}
The theoretical explanation of the ongoing accelerated expansion of the universe\cite{RiessAdamG:1998cb, WMAP:2003elm,WMAP:2006bqn,WMAP:2008lyn,SDSS:2003eyi,Jain:2003tba} is one of the most unsolved cosmological problems. Achieving accelerated expansion by adding a constant to Einstein-Hilbert(E-H) action suffers from fine-tuning problem\cite{Weinberg:1988cp}. There are many ways to explain the accelerated epoch of the current universe. 
In this context modified gravity theories have been developed by modifying the E-H action, for example, F(R) theories of gravity\cite{Capozziello:2002rd,Nojiri:2006gh,Sotiriou:2008rp,DeFelice:2010aj,Starobinsky:2007hu,Odintsov:2019evb}. 
In recent times another class of most popular modified gravity theories is nonlocal gravity models\cite{Dirian:2014ara, Belgacem:2020pdz}. These models are motivated by Einstein action's ultraviolet (UV) and infrared (IR) corrections. They also provide a theoretical explanation of the accelerated expansion of the current universe. Some salient features of nonlocal gravity models are: (1) they can be employed to study cosmology in both the infrared and the ultraviolet, (2) valid cosmological perturbation theory, and (3) the resulting cosmology has good agreement with most observations.

In this direction, the nonlocal terms with Ricci scalar $R$ and $F(R)$ are extensively studied along with Einstein Hilbert action. These terms involve analytic transcendental functions of the covariant d'Alembert operator $\Box$. Initially, as developed by Wetterich, models with correction terms like $ R\Box^{-1}R $ are shown to be effective IR corrected nonlocal gravity model\cite{Wetterich:1997bz}. Further, Deser and Woodard\cite{Deser:2007jk}  introduce a general form as$ Rf(\frac{1}{\Box}R) $ which can be responsible for the late-time cosmic expansion of the universe. In this line, for a recent review on this topic, refer to\cite{Capozziello:2022lic}. 

 The nonlocal theory can have an equivalent scalar-tensor form by introducing auxiliary variables. Higher derivative terms with d'Alembertian operator and curvature can be considered higher derivative field theories. According to Ostrogradsky theorem\cite{Ostrogradsky:1850fid}, non-degenerate higher derivative Lagrangians are cursed with instabilities (popularly known Ostrogradsky instabilities). Generally, these instabilities are easy to identify by linear momentum terms within the Hamiltonian, which make it unbounded above and below depending on the structure\cite{Woodard:2015zca}. However degenerate theories\cite{Horndeski:1974wa,Nicolis:2008in,Deffayet:2011gz,Kobayashi:2011nu,Gleyzes:2014dya,Gleyzes:2014qga,Langlois:2015cwa,Motohashi:2016ftl,Joshi:2019tzr,Joshi:2021azw}(for review, refer to \cite{Langlois:2018dxi,Kobayashi:2019hrl}) in this regard is free from these instabilities by reducing the phase space non trivially \cite{Chen:2012au}

 In order to check the appearance of Ostrogradsky instability in any higher derivative theory, a Hamiltonian analysis would be required. The Hamiltonian analysis of these types of theories can be performed by the Dirac method of constraint system\cite{Wipf:1993xg, Dirac:1958sq, Dirac:1958sc, dirac2013lectures, Matschull:1996up}. The Hamiltonian analysis of the nonlocal model with inverse powers of d'Alembertian operators acting only on  Ricci scalar is performed in\cite{Kluson:2011tb}. A more general analysis of constructing Hamiltonian in this context from a modern perspective is done in \cite{Joshi:2019cyk} including terms with inverse d'Alembertian operators acting on Riemann tensor, Ricci tensor, and Ricci scalar. 
 
 In this work, we extend the analysis of ref. \cite{Joshi:2019cyk} for a non local F(R) gravity models. Earlier works in this line can be found in\cite{DeFelice:2014kma, Nojiri:2019dio}. Authors of ref.\cite{DeFelice:2014kma} consider a class of nonlocal gravity in which the Lagrangian is a general function of $\Box^{-1}R$ and derive the condition for the appearance of ghost fields by considering the kinetic term of the localized Lagrangian.
 In another ref.  \cite{Nojiri:2019dio}, they have found the condition for the nonexistence of ghost degrees of freedom in nonlocal F(R) gravity. Here the ghosts are usually identified through the kinetic matrix of the Lagrangian with a negative determinant, and after appropriate modification in the Lagrangian, ghost-free condition can be obtained. Application of this model to cosmology has also been carried out in ref.\cite{Nojiri:2019dio}.
 
This work tries to shed some light on the results obtained in \cite{Nojiri:2019dio, DeFelice:2014kma} from the Hamiltonian point of view. Here we construct the Hamiltonian for the localized action and identify the linear momenta term in the Hamiltonian, which is usually responsible for the appearance of Ostrogradsky ghosts in any higher derivative theory. Then we show that by putting an appropriate condition in the structure of constraints, we can get rid of these ghosts analogous to the approach followed in \cite{Nojiri:2019dio, DeFelice:2014kma} by analyzing the kinetic term of the Lagrangian.  
For deriving Hamiltonian, first, we separate both spatial and time derivative terms in Lagrangian in terms of ADM variables, similar to the analysis done in \cite{Kluson:2011tb, Joshi:2019cyk}. Then we identify several constraints of our theory, which eventually lead us to count the number of ghost degrees of freedom.  
 
 This paper has many sections. In section II, ADM formalism is reviewed. In section III, we formulate the Hamiltonian for the F(R) nonlocal model and shed light on conditions that can give ghost-free theories. In section IV, we perform Hamiltonian formalism for a general action of $\Box^{-1}R$; we study the ghost structure from the Hamiltonian analysis. Finally, we summarise our results in section 5.

\section{ADM Formalism}
In this section, we review the 3 + 1 formalism's basic features in order to formulate the Hamiltonian analysis of the purposed theory.  Space-time is characterized by $(M,g_{\mu\nu} )$, where M is a 4-dimensional differentiable manifold and $g_{\mu\nu}$ is a Lorentzian metric.  $(M,g_{\mu\nu} )$ can be foliated by the family of space like surface ($\Sigma_{t}$), for more details refer to\cite{Arnowitt:1962hi,Gourgoulhon:2007ue,17b23f5522334b1ca8bb575ecaf5c01e}. Mathematically, four dimensional metric can be written in terms of induced metric $ h_{\mu\nu}$ and normal vector $ n_{\mu}$,
\begin{equation}
g_{\mu\nu}= h_{\mu\nu}- n_{\mu} n_{\nu}. \label{metric}
\end{equation}
The time-like future-directed vector $n_{\mu} $ is  normal to  the three dimensional space like surface, having a property,
$n_{\mu}n^{\mu}=-1$. The time direction vector $t^{a}=\frac{\partial}{\partial t}$, related to $n^{\mu}$ as,
\begin{equation}
t^{\mu}=Nn^{\mu}+N^{\mu},
\end{equation}
where, $N^{\mu}$ is the shift vector, and $N$ is the lapse function.
The line element in the  3+1 decomposition  reads  
\begin{equation}
ds^2= -N^{2}dt^{2}+h_{ij}(dx^{i}+N^{i}dt)(dx^{j}+N^{j}dt).
\end{equation} 
We used the indices $i,j....$ for showing spatial character of shift vector and spatial metric. The value of N and $ N^{i}$  in terms of metric is given as,
\begin{equation}
N=\frac{1}{\sqrt{-g^{00}}},\qquad N^i=-\frac{g^{0i}}{\sqrt{g^{00}}}.
\end{equation}
Further, the various metric elements can be written in terms of $h_{ij},N^{i}$ and $N$ as,
\begin{eqnarray}
g_{00}=-N^2+h_{ij}N^{i}N^{j},\qquad g_{0i}=N_{i},\qquad g_{ij}=h_{ij}, \nonumber\\ g^{00}=\frac{1}{N^{2}},\qquad g^{0i}=\frac{N^i}{N^2}, \qquad g^{ij}=h^{ij}-\frac{N^i N^j}{N^2}.
\end{eqnarray}
  The normal vector in terms of lapse and shift function are 
$$ n_{0}=-N, \qquad n_{i}=0, \qquad n^{0}=\frac{1}{N}, \qquad n^{i}=\frac{N^{i}}{N}.$$
The evolution of the spatial metric, $h_{ij}$ is given by the Lie derivative along $n^{a}$,
\begin{equation}
    \mathcal{L}_{n}h_{ij}=2NK_{ij}
\end{equation}
 where, $K_{ij}$ is the extrinsic curvature which is,
 \begin{eqnarray} 
K_{ij}=\frac{1}{2N}\left( \dot{h_{ij}}-\mathcal{D}_{i} N_{j}-\mathcal{D}_{j}N_{i}\right).
\end{eqnarray}
 Here the overdot represents the lie-derivative with respect to the time flow vector $t^{b}$ of the metric, and $\mathcal{D}_{i}$ denotes the spatial derivative.
 Evolution of $K_{ij}$ is related to $L_{n}K_{ij}$ which takes the form,
\begin{eqnarray}
\mathcal{L}_{n}K_{ij}=\frac{1}{N}(\dot{ K_{ij}}-\mathcal{L}_{\Vec{N}}K_{ij}).
\end{eqnarray}
where $\mathcal{L}_{\Vec{N}}$ is lie derivative with respect to shift vector, $\dot{ K_{ij}}$ contains second order time derivative of induced metric $h_{ij}$. The Ricci scalar in (3+1) decomposition takes the form,
\begin{equation}
R= \mathcal{R}+ K^{2}-3 K_{i j} K^{ij}+ 2 h^{ij} L_{n}  K_{ij} -\frac{2}{N}D_{i}D^{i}N. \label{RIS}
\end{equation}
Similarly $\mathcal{L}_{n} \phi$ containing the first order time derivative of any scalar  $\phi$ is defined as,
\begin{equation}
\mathcal{L}_{n}\phi=n^{a}\nabla_{a}\phi=\frac{1}{N}(\dot\phi-\mathcal{L}_{\Vec{N}}\phi). \label{DOS}
\end{equation}
The 3+1 decomposition of our action will help us construct canonical momenta for the Hamiltonian. Next, we derive a Hamiltonian for our model.
\section{Hamiltonian Formalism of Nonlocal F(R) Gravity model}
Let us consider action with nonlocal terms of the form\cite{Nojiri:2019dio},
\begin{equation}
S=\int d^{4}x \sqrt{-g}\left[G(R)+ F(R) \Box^{-k}H(R) \right], \label{action}
\end{equation}
where, $G(R)$, $ F(R) $ and $ H(R) $ are general functions of Ricci scalar, and  $k$ is a positive integer. This model is examined in detail for $k=1$ and $k=2$ from the Lagrangian point of view. It is shown that this theory is ghost free for $k=1.$ However it can be made to be free from ghosts by adding extra fields  for $k=2$. Next, our plan is to analyze the same model in  Hamiltonian formalism. For this we  write the above action in an equivalent  scalar-tensor form by introducing Lagrange multiplier C, which replaces R by Q,
 \begin{equation}
 S_{1}^{eqv} = \int d^{4}x  \sqrt{-g}\left[G(R)+ F(R) \Box^{-k}H(R)+ C(R-Q) \right],\label{s12}
\end{equation} 
 introducing different set of auxiliary fields $B_{k}$ and $A_{k}$\cite{Joshi:2019cyk}, 
 after some straightforward calculation, eq.(\ref{s12}) can be  written as,
\begin{align}
S^{eqv} = \int d^{4}x\sqrt{-g} \bigg[ G(Q) + F(Q)A_{1} + C\left( R -Q\right) -B_{k}H(Q)  \label{ass}\\\nonumber -\sum_{n=1}^{k-1} B_{n} A_{n+1}+\sum_{n=1}^{k} B_{n}\Box A_{n} \bigg].
\end{align}
The last term of action eq.(\ref{ass}) simplifies to
\begin{equation}
\int d^{4}x N\sqrt{h}\sum_{L=1}^{\infty} B_{n }\Box A_{n}=-\int d^{4}x  N\sqrt{h}\sum_{L=1}^{\infty}\nabla _{\rho}B_{n} \nabla^{\rho} A_{n}+\textnormal{surface term}.\label{eq:ss1}
\end{equation}
Further, the  eq.(\ref{ass}) in terms of ADM variables looks,
\begin{eqnarray}
\begin{aligned}
S^{eqv}= \int d^{3}x N\sqrt{h} \bigg [G(Q) +  F(Q)A_{1} + C\left( K_{ij}K^{ij} -K^{2} + \mathcal{R} -Q\right) &\\ -2K \ (n^{a} \nabla_{a}C)  -  \sum_{n= 1}^{k-1}  B_{n} A_{n+1} - B_{k}H(Q) - h^{ij} \sum_{n=1}^{k}  \mathcal{D}_{i}B_{n}\mathcal{D}_{j}A_{n} &\\+ \sum_{n=1}^{k}  (n^{a}\nabla_{a}B_{n})(n^{b}\nabla_{b}A_{n}) +\mathcal{D}_{i}\mathcal{D}^{i}C\bigg].\label{ADMaction}
\end{aligned}
\end{eqnarray}
Now we derive the canonical momenta  with respect to set of variables $(N, N^i, h_{ij}, Q, A_{1},A_{2}...A_{n}, B_{1},B_{2}...B_{n})$, which are 
\begin{eqnarray}
\begin{aligned}
\Pi_{N} & \approx 0, \ \ \ \ \ 
\Pi_{i} \approx 0,\ \ \ 
\Pi^{ij} = \sqrt{h} C( \mathcal{K}^{ij} -h^{ij}\mathcal{K} ) - \sqrt{h} h^{ij} (n^{a} \nabla_{a}C), \qquad \qquad \qquad \qquad   \\
P_{Q} & \approx  0 , \qquad \qquad
P_{C} = -2\sqrt{h} \mathcal{K}, \qquad\\
P_{A_{1}} &= \sqrt{h} (n^{a}\nabla_{a}B_{1}), \ \ P_{A_{2}} = (n^{a}\nabla_{a}B_{2}) \  ,,,,,,,\ P_{A_{k}}=\sqrt{h} (n^{a}\nabla_{a}B_{k})
\\ P_{B_{1}}& = \sqrt{h}  (n^{a}\nabla_{a}A_{1})\  \ P_{B_{2}} = \sqrt{h}  (n^{a}\nabla_{a}A_{2}),,,,,,,,P_{B_{k}} = \sqrt{h}  (n^{a}\nabla_{a}A_{k}).\label{cmo}
\end{aligned}
\end{eqnarray}
Here we use $\Pi$ notation for denoting momenta for variables appearing from the gravity side $(N, N^i, h_{ij})$ and P for other auxiliary variables $(C, A_{1},A_{2}...A_{n}, B_{1},B_{2}...B_{n}).$ For further calculations, we follow the Dirac method for deriving the Hamiltonian for constraints system, in which variables with vanishing canonical momenta as primary constraints are denoted by the $\approx$ sign in the constraint space $\Gamma$.
Thus, from the eq.(\ref{cmo}), the primary constraints in our theory are,
\begin{equation*}
\Pi_{N}\approx 0 ,\ \ \Pi_{i}\approx 0 ,\ \ P_{Q}\approx 0.
\end{equation*}
After deriving all the canonical momenta and identifying primary constraints, we can write Hamiltonian density for the corresponding Lagrangian eq.(\ref{ADMaction}) as,
\begin{eqnarray}
\mathcal{H}=\Pi^{ij}\dot{h}_{ij}+p^C \dot{C} +\sum_{n = 1}^{k}( P_{B_{n}}\dot{B}_{n} + P_{A_{n}} \dot{A}_{n})-L
\end{eqnarray}
The time derivative of all variables to  their corresponding momenta can be extracted form eq.(\ref{cmo}).
After putting these values, the final  form of Hamiltonian  can be written as,
\begin{equation}
\mathcal{H}= N\mathcal{H}_{N}  + N^{i} \mathcal{H}_{i}, 
\end{equation}
with
\begin{eqnarray}
\begin{aligned}
\mathcal{H}_{N} & = \frac{1}{\sqrt{h} C} \Pi^{ij} h_{ik} h_{jl} \Pi^{kl} -\frac{1}{3\sqrt{h} C} \Pi^{2} -\frac{1}{3\sqrt{h}} \Pi P_{C} + \frac{1}{6 \sqrt{h}} C P_{C}^{2} -\sqrt{h} C \mathcal{R}+ \sqrt{h}CQ  \\& - \sqrt{h}  G(Q) - F(Q)A_{1} +\sqrt{h}\sum_{n=1}^{k} A_{n+1}B_{n} +\sqrt{h}B_{k}H(Q) \\&  +\sum_{n = 1}^{k} \frac{P_{A_{n}}P_{B_{n}}}{\sqrt{h}} -\sqrt{h} h^{ij} \sum_{n = 1}^{k} \Big( \mathcal{D}_{i}B_{n} \mathcal{D}_{j}A_{n}\Big),\label{eq201} 
\end{aligned}
\end{eqnarray}
and
\begin{equation}
\begin{split}
\mathcal{H}_{i}  = -2 h_{ik}\mathcal{D}_{j}\Pi^{kj} + \sum_{n = 1}^{k} \Big(P_{A_{n}}\mathcal{D}_{i}A_{n} + P_{B_{n}}\mathcal{D}_{i}B_{n} \Big) + P_{C}\mathcal{D}_{i}C.
\end{split}
\end{equation} 
The $\Pi$ is defined as, $h_{ij}\Pi^{ij}=\Pi$.  The total Hamiltonian density with primary constraints in terms of Lagrange multipliers $ \lambda^{i},\lambda^{N} $ and $\lambda^{Q}$ takes the form,
\begin{equation}
\mathcal{H}_{tot} =  N\mathcal{H}_{N} + N^{i}\mathcal{H}_{i} +\lambda^{i}\Pi_{i}+\lambda^{N}\Pi_{N} + \lambda^{Q}P_{Q},
\end{equation}
and the total Hamiltonian is expressed as,
\begin{equation}
H_{total} = \int d^{3}x \Big( N \mathcal{H}_{N}  + N^{i} \mathcal{H}_{i} +\lambda^{i}\Pi_{i}+\lambda^{N}\Pi_{N} + \lambda^{Q}P_{Q} \Big),\label{eq;TH}
\end{equation}
 In Dirac method for constraints system, the constraints are classified according to their nature of time evolution. For $\Pi_{N}$ and $\Pi_{i}$, we get
\begin{eqnarray}
\dot{\Pi_{N}}=\{\Pi_{N},\mathcal{H}_{tot}\}=H_N,~~ \qquad \dot{\Pi_{i}}=\{\Pi_{i},\mathcal{H}_{tot}\}=H_i,
\end{eqnarray}
where $H_{N}$ and $H_{i}$ are secondary constraints, are known as Hamiltonian and diffeomorphism constraints respectively. On the constraint space $ \Gamma, $ the time evolution of constraints $H_N$ and $H_i$ vanishes weakly as,
\begin{equation}
\dot{\mathcal{H_N}}=\{\mathcal{H}_N,H_{tot}\}\approx0, \label{hn}
\end{equation} 
and
   \begin{equation}
\dot{\mathcal{H}}_i=\{\mathcal{H}_i,H_{tot}\}\approx0. \label{hi}
    \end{equation} 
Further, we evaluate time evolution of  primary constraint $P_Q$ with total Hamiltonian, 
\begin{eqnarray}
\Xi_{Q} = \partial_{t}P_{Q} =\{ P_{Q}, \mathcal{H}_{tot}\} = N\sqrt{h}\Big\{-C +G'(Q) +F'(Q)A_{1} - H'(Q)B_{k}\Big\}\approx 0. \label{con1}
\end{eqnarray}
where, the dashed sign (') denotes the derivative with respect to Q, for example $G'(Q)=\frac{\partial G(Q)}{\partial Q}$. $\Xi_{Q}\approx 0$, $\Xi_{Q}$ act as secondary constraint corresponding to primary constraint $P_Q\approx 0$.  Now, we check time evolution of $\Xi_{Q}$ and demanding that it vanishes on constraint surface we obtain,
\begin{eqnarray}
\begin{aligned}
\dot{\Xi}_{Q} =& \{ \Xi_{Q},\mathcal{H}_{tot}\}=
N \Bigg[ \frac{1}{3}(\Pi -CP_{C}) + F'(Q)P_{B_{1}} 
 -H'(Q)P_{A_{k}}\Bigg] \\& + \sqrt{h}\lambda^{Q}(G''(Q) +F''(Q)A_{1}-H''(Q)B_{k})\approx0. \label{aab}
 \end{aligned}
\end{eqnarray}
where the double dash (' ') show second order derivative with respect to $Q$.
This condition eq.(\ref{aab}) fixes the Lagrange multiplier $\lambda^{Q}$, and hence no tertiary constraint appears in this theory.\\  Now, we have identified all the primary and secondary constraints. Then we categorize
them into first and second-class constraints. For which, we  check the  Poisson bracket of  $P_Q$ and $\Xi_{Q}$,
\begin{equation}
\{P_Q,\Xi_{Q}\}\neq 0,
\end{equation}
and it implies that $P_Q$ and $\Xi_{Q}$ are second class constraints.
All other Poisson brackets are obtained as,
\begin{eqnarray}
\{ \Pi_{N}, \Pi_{i}\}= \{ \Pi_{N}, \Pi_{N}\} = \{ \Pi_{N}, \mathcal{H}_{N}\}=\{ \Pi_{N},\mathcal{H}_{i}\} \approx 0 ,\nonumber \\
 \{ \Pi_{i}, \Pi_{i}\} = \{ \Pi_{i},\mathcal{H}_{N}\} = \{ \Pi_{i},\mathcal{H}_{i}\} \approx 0 ,\nonumber \\
 \{ P_{Q}, P_{Q}\} = \{ P_{Q}, \Pi_{N}\} = \{ P_{Q}, \Pi_{i}\} \approx 0.
 \end{eqnarray}
In summary, we have  eight first class constraints $(\Pi_{N}, \Pi_{i}, \mathcal{H}_{N}, \mathcal{H}_{i})$ and two second class constraints $P_Q $ and $ G_Q.$
Now we can derive results for various choices of general function  $G(R)$, $ F(R) $, and $ H(R) $  from the above analysis.
\subsection{Analysis for Different Choices of \texorpdfstring{$G(R)$, $ F(R) $}{TEXT} and \texorpdfstring{$ H(R) $}{TEXT}}
Here we reiterate that the Ostrogradsky ghost is seen in the Hamiltonian by linear momenta terms. It can be observed from eq.(\ref{eq201}) that there are k terms of linear momenta of auxiliary variables $A_{k}$  and $B_{k}$. We also note that for the general function  $G(R)$, $ F(R) $, and $ H(R) $,  we obtain only two constraint equations. These constraints are not sufficient to remove all the linear momenta terms from eq.(\ref{eq201}). Now we analyze different cases.
\begin{itemize}
    \item For all our analysis we set  $C=G'(Q)$ in eq.(\ref{con1}. Then we obtain the form of constraint $\dot {\Xi}_{Q}$ as,
\begin{eqnarray}
\Xi_{Q}^{II}=\dot{\Xi}_{Q} =
N \Bigg[ F'(Q)P_{B_{1}} 
 -H'(Q)P_{A_{k}}\Bigg] + \sqrt{h}\lambda^{Q}(F''(Q)A_{1}-H''(Q)B_{k})\approx0, \label{aab10}
\end{eqnarray}
\end{itemize}

\textbf{Case1}: If we choose $ F''(Q)=0$ and $ H''(Q)=0$, then the Lagrange multipliers remain undetermined. And these conditions hold only if $ F(Q)$ and $ H(Q)$ are linear functions of $Q,$ which provides a relation between
$ F(Q)$ and $ H(Q)$ i.e $ H(Q)=\alpha F(Q)+\textnormal{any constant}.$ For this case the eq. \ref{aab10} reads
\begin{eqnarray}
\Xi_{Q}^{II}=\dot{\Xi}_{Q}^{} = N \Bigg[  P_{B_{1}} -\alpha P_{A_{k}}\Bigg]\approx0.
\end{eqnarray}
 Due to undetermined nature of Lagrange multiplier, the time evolution  $\Xi_{Q}^{II}$ with Hamiltonian generate new constraint (tertiary constraint),
\begin{eqnarray}
\Xi_{Q}^{2I}=\dot{\Xi}_{Q}^{II} = N \Bigg[  A_{2} -\alpha B_{k-1}\Bigg]\approx0,
\end{eqnarray}
and the time evolution  $\Xi_{Q}^{2I}$ with Hamiltonian becomes,
\begin{eqnarray}
\Xi_{Q}^{2II}=\dot{\Xi}_{Q}^{} = N \Bigg[  P_{B_{2}} -\alpha P_{A_{k-1}}\Bigg]\approx0.
\end{eqnarray}
 Next we repeat the above procedure for n times which yields,
\begin{eqnarray}
\Xi_{Q}^{nI}=\dot{\Xi}_{Q}^{n-1} = N \Bigg[  A_{n} -\alpha B_{1}\Bigg]\approx0.\label{eq:op}
\end{eqnarray}
 Time evolution of  $\Xi_{Q}^{nI}$ eq.(\ref{eq:op}) becomes,
\begin{eqnarray}
\Xi_{Q}^{nII}=\dot{\Xi}_{Q}^{I} = N \Bigg[  P_{B_{n}} -\alpha P_{A_{1}}\Bigg]\approx0.
\end{eqnarray}
Now the chain of constraints ends and we have total $2n$ constraints  ($\Xi_{Q}^{},\Xi_{Q}^{II},\ \Xi_{Q}^{2I},\Xi_{Q}^{2II},\ .....,\ \Xi_{Q}^{nI},\Xi_{Q}^{nII}$).
These constraints relate $A_{1}$ to $ B_{n}$, $A_{2}$ to $B_{n-1}$...$A_{n}$ to $B_{1}$, and similarly relate momenta $ P_{B_{n}}$  to $ P_{A_{1}}$,  $P_{B_{n-1}}$ to $P_{A_{2}}$.....$ P_{B_{1}}$  to $ P_{A_{n}}.$ However for removing the linear momenta terms in Hamiltonian eq.(\ref{eq201}) relations between different momenta, i.e., $ P_{B_{1}}$  to $ P_{A_{1}}$,  $P_{B_{2}}$ to $P_{A_{2}}$.....$ P_{B_{n}}$  to $ P_{A_{n}}$ are required. Because of the absence of these one to one relation in this theory, we get $n$ ghosts.\\
For a special case like $k=1$, eq.(\ref{con1}) becomes, 
\begin{eqnarray}
\Xi_{Q}=N\sqrt{h}\big\{\alpha A_{1}-B_{1}\big\}\approx0,\label{se:eq}
\end{eqnarray}
and its time evolution is,
 \begin{eqnarray}
\Xi_{Q}^{II}=\dot{\Xi}_{Q}^{} = N \Bigg[  P_{B_{1}} -\alpha P_{A_{1}}\Bigg]\approx0.\label{se1:eq}
\end{eqnarray} 
 Eq.\ref{se1:eq}  relates $P_{A_{1}}$ to $P_{B_{1}}$. Consequently, $H_{N}$ in eq.(\ref{eq;TH}) reads,
\begin{eqnarray}
\begin{aligned}
\mathcal{H}_{N} & = \frac{1}{\sqrt{h} C} \Pi^{ij} h_{ik} h_{jl} \Pi^{kl} -\frac{1}{3\sqrt{h} C} \Pi^{2}  -\sqrt{h} C \mathcal{R}+ \sqrt{h}CQ  \\& - \sqrt{h}  G(Q) + \frac{P_{A_{1}}^{2}}{\sqrt{h}} -\sqrt{h} h^{ij} \sum_{n = 1}^{k} \Big( \mathcal{D}_{i}A_{1} \mathcal{D}_{j}A_{1}\Big),
\end{aligned}
\end{eqnarray}
Here all the momenta terms are quadratic, and there is no place for Ostrogradsky ghost to appear.

\begin{itemize}
    \item \textbf{Case2}: Here we choose $ H(Q)=Q$ and $ F(Q)$ is not a linear function of Q. In this case then the  form of $\dot {\Xi}_{Q}$ is,
\begin{eqnarray}
\Xi_{Q}^{II}=\dot{\Xi}_{Q} = N \Bigg[ F'(Q) P_{B_{1}} -P_{A_{k}}\Bigg]  + \sqrt{h}\lambda^{Q} F''(Q)A_{1}\approx0.\label{eq27}
\end{eqnarray}
It can be observed from eq.(\ref{eq27}) that by fixing the Lagrange multiplier, we cannot get rid of all the linear momenta terms in the Hamiltonian. A model of this kind is formulated in\cite{Nojiri:2019dio} from the Lagrangian point of view, where authors draw the result that ghosts exit as the determinant of kinetic matrix takes negative values. The same conclusions from our Hamiltonian analysis for this particular model can be drawn here. \\

 \textbf{Case3}: Here we take $ F(Q)$=$ H(Q)$ for which the structure of $\Xi_{Q}^{II}$ becomes
    \begin{eqnarray}
    \Xi_{Q}^{II}=\dot{\Xi}_{Q} = NF'(Q) \Bigg[ P_{B_{1}} -P_{A_{k}}\Bigg] + \sqrt{h}\lambda^{Q}F''(Q)( A_{1}-B_{k})\approx0.\label{eq:aag}
    \end{eqnarray}
    In this case we can fix the Lagrange multiplier which in turn provides only two constraints. However these constraints are not sufficient to remove all the linear momenta term from the Hamiltonian eq.(\ref{eq201}).
    
    Now, the theory with $k=1$, we can remove the linear momenta term in the Hamiltonian by imposing the condition that $A_{1}$ and $B_{1}$ are  related to each other linearly i.e. $A_{1}=p B_{1},$ (p=constant). Under this condition no ghosts appear for $k=1$ case and the constraint eq.(\ref{con1}) becomes,
    \begin{eqnarray}
    \Xi_{Q} = \partial_{t}P_{Q} =\{ P_{Q}, \mathcal{H}_{tot}\} = N\sqrt{h}\Big\{F'(Q)A_{1}( 1- p)\Big\}\approx 0,
    \end{eqnarray}
    and its time evolution with Hamiltonian reads,
 \begin{eqnarray}
    \Xi_{Q}^{II}=\dot{\Xi}_{Q} = NF'(Q) P_{B_{1}}\Bigg[ 1-p \Bigg]+\sqrt{h}\lambda^{Q}F''(Q)A_{1}(1-p)\approx0.
    \end{eqnarray}
    
 For a particular value of $p$, i.e., $p=1,$ then $A_{1}= B_{1}$ and their corresponding momenta are equal $ P_{A_{1}} = P_{B_{1}}.$ This implies $ \Xi_{Q}=0.$ This is a exact relation and happens only for special case when $p=1.$ However for other values of $p$ we have to use usual procedure to show the nonexistence of ghosts. Here we conclude that for $k=1$ no Ostrogradsky ghost is present for the condition when auxiliary variables depend linearly on each other .
\end{itemize}
Next we consider $k=2$ and the structure of constraints takes the form,
\begin{eqnarray}
\Xi_{Q} = \partial_{t}P_{Q} =\{ P_{Q}, \mathcal{H}_{tot}\} = N\sqrt{h}\Big\{ F'(Q)(A_{1} -B_{2})\Big\}\approx 0.\label{eq:cca}
\end{eqnarray}
and  its evolution gives,
\begin{eqnarray}
\begin{aligned}
\dot{\Xi}_{Q} =& \{ \Xi_{Q},\mathcal{H}_{tot}\}=
N \Bigg[ F'(Q)(P_{B_{1}} 
 -P_{A_{2}})\Bigg] + \sqrt{h}\lambda^{Q}(F''(Q)(A_{1}-B_{2})\approx0. \label{aac}
 \end{aligned}
\end{eqnarray}
Similar to the the analysis performed for $k=1$, the mathematical structure of constraints eq.(\ref{eq:cca}) and  eq.(\ref{aac} relate  $A_{1} $ to $B_{2}$, and $ P_{B_{1}}$ to $P_{A_{2}}$. On top of it a linear relation between $ P_{B_{1}}$ and $P_{A_{1}}$ is required to make the theory ghost free. But there is no way to achieve this for $k=2.$  Generalizing the above analysis for values of $k\geq 2,$ it can be shown that ghosts always appear.   

For $k=2$, \cite{Nojiri:2019dio}, suggested a method for developing a ghost-free theory, we need to modify the Lagrangian. For clarity, we again reformulate the Hamiltonian formalism for $k=2$. Now, we start with an action\cite{Nojiri:2019dio},
 \begin{equation}
S=\int d^{4}x \sqrt{-g}\left[R-\frac{1}{2} F(R) \Box^{-2}F(R) \right].\label{eq:aat}
\end{equation}
 Following \cite{Nojiri:2019dio}, after introducing  $B_{1}=\frac{1}{2}\tilde{A}+\tilde{B}$ and  $B_{2}=\tilde{A}-2\tilde{B}$ in eq.(\ref{ass}), the  final scalar tensor form of eq.(\ref{eq:aat}) reads,
\begin{align}
S^{eqv} = \int d^{4}x\sqrt{-g} \bigg[ Q - F(Q)A_{1} + C\left( R -Q\right) -  \tilde{A}^2+A_{1}\Box \tilde{A} \bigg].
\end{align}
Despite these procedures, the Hamiltonian may still include linear momenta terms due to the presence of the term $A_{1}\Box \tilde{A}.$  Next, we adopt a method put forward by authors of ref.\cite{Nojiri:2019dio} where the kinetic term is modified to obtain a ghost-free action. Modified action takes the form,
\begin{eqnarray}
S_{mod} = S^{eqv}+ \int d^{4}x\sqrt{-g} \bigg[\frac{\beta}{2}A_{1}\Box A_{1}+\frac{\alpha}{2}\tilde{A}\Box \tilde{A} \bigg],
\end{eqnarray}
and after putting $S^{eqv}$,
\begin{align}
S_{mod} = \int d^{4}x\sqrt{-g} \bigg[ Q - F(Q)A_{1} + C\left( R -Q\right) -  \tilde{A}^2+A_{1}\Box \tilde{A}+\frac{\beta}{2}A_{1}\Box A_{1}+\frac{\alpha}{2}\tilde{A}\Box \tilde{A}].\label{q;mid}
\end{align}
 3+1 decomposition of $S_{mod}$ becomes, 
\begin{eqnarray}
\begin{aligned}
S_{mod} =& \int d^{4}x\sqrt{-g} \bigg[ Q -F(Q)A_{1} + C\left( K_{ij}K^{ij} -K^{2} + \mathcal{R} -Q\right)  -2K \ (n^{a} \nabla_{a}C)-  \tilde{A}^2\\&+(n^{a}\nabla_{a}A_{1})(n^{b}\nabla_{b}\tilde{A})+\frac{\beta}{2}(n^{a}\nabla_{a}A_{1})(n^{b}\nabla_{b}A_{1})+\frac{\alpha}{2}(n^{a}\nabla_{a}\tilde{A})(n^{b}\nabla_{b}\tilde{A})- h^{ij}\mathcal{D}_{i}A_{1}\mathcal{D}_{j}\tilde{A} \\&-\frac{\beta}{2}h^{ij}\mathcal{D}_{i}A_{1}\mathcal{D}_{j}A_{}-\frac{\alpha}{2}h^{ij}\mathcal{D}_{i}\tilde{A}\mathcal{D}_{j}\tilde{A}+\mathcal{D}_{i}\mathcal{D}^{i}C\bigg].
\end{aligned}
\end{eqnarray}
Here, we show the canonical momenta related to $A_{1}$ and $\tilde{A}$ only,  and the canonical momenta relation for $C, N, N^{i}$ can be calculated similarly to the procedure used for previous cases of study. Thus,
\begin{eqnarray}
P_{\tilde{A}}=\frac{\partial L}{\partial \tilde{A}}=\sqrt{h}\left\{\left(n^{a}\nabla_{a}A_{1}\right)+\alpha \left(n^{a}\nabla_{a}\tilde{A}\right)\right\}\nonumber\\P_{A_{1}}=\frac{\partial L}{\partial A_{1}}=\sqrt{h}\left\{(n^{a}\nabla_{a}\tilde{A})+\beta (n^{a}\nabla_{a}A_{1})\right\}
\end{eqnarray}
After solving these equations for the $n^{a}\nabla_{a}\tilde{A}$ and $n^{a}\nabla_{a}A_{1}$ is, 
\begin{eqnarray}
n^{a}\nabla_{a}\tilde{A}=\frac{1}{\sqrt{h}(\alpha\beta-1)}(\beta P_{\tilde{A}}- P_{A_{1}})\\
n^{a}\nabla_{a}A_{1}=\frac{1}{\sqrt{h}(\alpha\beta-1)}(\alpha P_{A_{1}}- P_{\tilde{A}})
\end{eqnarray}
The Hamiltonian of action eq.(\ref{q;mid}) is written as, 
\begin{eqnarray}
\mathcal{H}=\Pi^{ij}\dot{h}_{ij}+p^C \dot{C} + P_{\tilde{A}} \dot{\tilde{A}}+ P_{A_{1}} \dot{A}_{1}-L.
\end{eqnarray}
After some simplification $\mathcal{H}$ is
\begin{equation}
\mathcal{H}= N\mathcal{H}_{N}  + N^{i} \mathcal{H}_{i}, 
\end{equation}
with
\begin{eqnarray}
\begin{aligned}
\mathcal{H}_{N} & = \frac{1}{\sqrt{h} C} \Pi^{ij} h_{ik} h_{jl} \Pi^{kl} -\frac{1}{3\sqrt{h} C} \Pi^{2} -Q + F(Q)A_{1} +  \tilde{A}^2+ h^{ij}\mathcal{D}_{i}A_{1}\mathcal{D}_{j}\tilde{A} \\&+\frac{\beta}{2}h^{ij}\mathcal{D}_{i}A_{1}\mathcal{D}_{j}A_{1}+\frac{\alpha}{2}h^{ij}\mathcal{D}_{i}\tilde{A}\mathcal{D}_{j}\tilde{A}+\frac{1}{2\,\alpha\,\beta-2\,}(\beta P_{\tilde{A}}^{2}-2\,P_{\tilde{A}}\,P_{A_{1}}+\alpha P_{A_{1}}^{2})
\end{aligned}
\end{eqnarray}
and
\begin{equation}
\begin{split}
\mathcal{H}_{i}  = -2 h_{ik}\mathcal{D}_{j}\Pi^{kj} +  \Big(P_{\tilde{A}}\mathcal{D}_{i}\tilde{A} + P_{A_{1}}\mathcal{D}_{i}A_{1} \Big) + P_{C}\mathcal{D}_{i}C.
\end{split}
\end{equation} 

As we can see from above Hamiltonian that linear terms of $P_{\tilde{A}}$ and $P_{A_{1}}$ appear. We can remove the linear momenta terms only if $\alpha\beta=1$, but the Hamiltonian is not finite with this choice. Therefore, the above method is not effective in removing linear terms of momenta from the Hamiltonian. In other words, a Hamiltonian formalism for $\alpha\beta=1$ is unphysical. However, we will now show that it is possible to find the viable parameter space where no ghosts appear after diagonalization. To do this, let us start with the kinetic part of the last term of $H_{N}$, which can take the form, 
\begin{gather}
 \begin{bmatrix} P_{\tilde{A}} & P_{A_{1}}  \end{bmatrix}
  \begin{bmatrix}
  \alpha  & -1 \\
  -1 & \beta 
   \end{bmatrix}
   \begin{bmatrix}
   P_{\tilde{A}} \\ P_{A_{1}}
   \end{bmatrix}
\end{gather}
Let $M= \begin{bmatrix}
  \alpha  & -1 \\
  -1 & \beta 
   \end{bmatrix}$ and after diagnolisation, we get the kinetic matrix as
   \begin{equation}
 \left[ \begin {array}{cc} \beta/2+\alpha/2-1/2\,\sqrt {{\alpha}^{2}-2
\,\beta\,\alpha+{\beta}^{2}+4}&0\\ \noalign{\medskip}0&\beta/2+\alpha/
2+1/2\,\sqrt {{\alpha}^{2}-2\,\beta\,\alpha+{\beta}^{2}+4}\end {array}
 \right]
\end{equation}
Both the diagonal term should be positive for non-existence of ghosts we obtain the following conditions
\begin{eqnarray}
 \beta+\alpha> \sqrt {{\alpha}^{2}-2
\,\beta\,\alpha+{\beta}^{2}+4},
\end{eqnarray} 
which holds only if
\begin{eqnarray}
\,\alpha\,\beta>1, \, \, \textnormal{and}\, \, \alpha+\beta>0.
\end{eqnarray} This matches with the results obtained in \cite{Nojiri:2019dio}.
Next, we will create a Hamiltonian for some generalized nonlocal gravity models.
\section{Hamiltonian Formalism for \texorpdfstring{$f(R,\Box^{-1}R,\Box^{-2}R...\Box^{-n}R)$}{TEXT} Gravity Model}
In this section, our aim is to perform Hamiltonian analysis  of the   model purposed in\cite{DeFelice:2014kma}. Here Lagrangian has a form as
\begin{eqnarray}
L=\sqrt{-g}\ f(R,\Box^{-1}R,\Box^{-2}R..........\Box^{-n}R),\label{ac4:eq}
\end{eqnarray}
where, f is general function of $\Box^{-n}R$ $(n=1,2,3...)$ and $n$ is finite. Then action eq.(\ref{ac4:eq})  satisfies  the following condition,
 \begin{equation}
     \frac{\partial L}{\partial R \partial (\Box^{-n}R})\neq 0.
 \end{equation}
By defining auxiliary variables $Q, A_{1},..A_{n}, B_{1},...B_{n}, C,$ the above non local action eq.(\ref{ac4:eq}) can be written in localized form,
\begin{eqnarray}
L=\sqrt{-g}\bigg[ f(Q,A_{1},A_{2},.......A_{n})+C(R-Q)+B_{1}(Q-\Box A_{1})\\\nonumber+B_{2}(A_{1}-\Box A_{2}),.........B_{n}(A_{n-1}-\Box A_{n})\bigg],
\end{eqnarray}
and further with straightforward calculation, transformed Lagrangian becomes,
\begin{eqnarray}
L=\sqrt{-g}\bigg[ f(Q,A_{1},A_{2},.......A_{n})+CR+Q( B_{1}-C)+g^{\mu\nu}(\partial_{\mu}B_{1}\partial_{\nu} A_{1}+\\\partial_{\mu}B_{2}\partial_{\nu}A_{2}......... \partial_{\mu}B_{n}\partial_{\nu}A_{n})\nonumber+B_{2}A_{1} +B_{3}A_{2}........B_{n}A_{n-1}\bigg].\label{red:eq}
\end{eqnarray}
Using  3+1 decomposition the eq.(\ref{red:eq}) reads,
\begin{eqnarray}
\begin{aligned}
S^{eqv}= \int d^{3}x N\sqrt{h} \bigg [f(Q,A_{1},A_{2},.......A_{n})  + C\left( K_{ij}K^{ij} -K^{2} + \mathcal{R} -Q\right) &\\ -2K \ (n^{a} \nabla_{a}C)+Q(B_{1}-C)+\mathcal{D}_{i}\mathcal{D}^{i}C  -  \sum_{k= 1}^{n-1}  B_{k} A_{k+1} &\\ + h^{ij} \sum_{n=1}^{k}  \mathcal{D}_{i}B_{n}\mathcal{D}_{j}A_{n} - \sum_{n=1}^{k}  (n^{a}\nabla_{a}B_{n})(n^{b}\nabla_{b}A_{n}) \bigg]
\end{aligned}
\end{eqnarray}
Likewise here the canonical momenta with respect to set of variables  $(N, N^i, h_{ij}, Q, C, A_{L}, B_{L})$ are defined by the following relations, 
\begin{eqnarray}
\begin{aligned}
\Pi_{N} & \approx 0, \ \ \ \ \ 
\Pi_{i} \approx 0,\ \ \ 
\Pi^{ij} = \sqrt{h} C( \mathcal{K}^{ij} -h^{ij}\mathcal{K} ) - \sqrt{h} h^{ij} (n^{a} \nabla_{a}C), \qquad \qquad \qquad \qquad   \\
P_{Q} & \approx  0 , \qquad \qquad
P_{\Phi} = -2\sqrt{h} \mathcal{K}, \qquad\\
P_{A_{1}} &= \sqrt{h} (n^{a}\nabla_{a}B_{1}), \ \ P_{A_{2}} = (n^{a}\nabla_{a}B_{2}) \  ,,,,,,,\ P_{A_{n}}=\sqrt{h} (n^{a}\nabla_{a}B_{n})
\\ P_{B_{1}}& = \sqrt{h}  (n^{a}\nabla_{a}A_{1})\  \ P_{B_{2}} = \sqrt{h}  (n^{a}\nabla_{a}A_{2}),,,,,,,,P_{B_{n}} = \sqrt{h}  (n^{a}\nabla_{a}A_{n}).\label{Con:eq}
\end{aligned}
\end{eqnarray}
Here, the primary constraints are $\Pi_{N} \approx 0,\Pi_{i} \approx 0, P_{Q}\approx 0$. The final form of Hamiltonian is given as,
\begin{equation}
\mathcal{H}= N\mathcal{H}_{N}  + N^{i} \mathcal{H}_{i,} 
\end{equation}
with
\begin{eqnarray}
\begin{aligned}
\mathcal{H}_{N} & = \frac{1}{\sqrt{h} C} \Pi^{ij} h_{ik} h_{jl} \Pi^{kl} -\frac{1}{3\sqrt{h} C} \Pi^{2} -\frac{1}{3\sqrt{h}} \Pi P_{C} + \frac{1}{6 \sqrt{h}} C P_{C}^{2} -\sqrt{h} C \mathcal{R} \\& - \sqrt{h}  f(Q,A_{1},A_{2},.......A_{n})-Q(B_{1}-C)-\mathcal{D}_{i}\mathcal{D}^{i}C\\&  +  \sum_{k= 1}^{n-1}  B_{k} A_{k+1}  - h^{ij} \sum_{n=1}^{k}  \mathcal{D}_{i}B_{n}\mathcal{D}_{j}A_{n} + \sum_{n=1}^{k}  P_{B_{n}}P_{A_{n}}, \label{HN:eq}
\end{aligned}
\end{eqnarray}
and
\begin{equation}
\begin{split}
\mathcal{H}_{i}  = -2 h_{ik}\mathcal{D}_{j}\Pi^{kj} + \sum_{n = 1}^{k} \Big(P_{A_{n}}\mathcal{D}_{i}A_{n} + P_{B_{n}}\mathcal{D}_{i}B_{n} \Big) + P_{\Phi}\mathcal{D}_{i}\Phi.
\end{split}
\end{equation} 
The behaviour of primary constraints $\Pi_{N} \approx 0,
\Pi_{i} \approx 0$ is analogous to analysis discussed in \textsection III. Here we mainly emphasize on  $P_{Q}$ and its evolution with total Hamiltonian gives rise to,
\begin{eqnarray}
\Xi_{Q} = \partial_{t}P_{Q} =\{ P_{Q}, \mathcal{H}_{tot}\} = N\sqrt{h}\Big\{B_{1}-C -f'\Big\}\approx 0. \label{CE:eq}
\end{eqnarray} 
where $f'$ denotes derivative of $f$ with respect to to Q. Further the  evolution of $\Xi_{Q}$ yields,
\begin{eqnarray}
\Xi_{Q}^{II}=\dot{\Xi}_{Q} = N \Bigg[\frac{1}{3}(\Pi-C P_{C}) +P_{A_{1}} -\frac{\partial f'}{\partial A_{1}}P_{B_{1}}+\frac{\partial f'}{\partial A_{2}}P_{B_{2}}+....\frac{\partial f'}{\partial A_{j}}P_{B_{j}} \\\nonumber+.... \frac{\partial f'}{\partial A_{n}}P_{B_{n}}\Bigg]  + \sqrt{h}\lambda^{Q}( f'')\approx0.\label{CE2:eq}
\end{eqnarray}
Using $\Xi_{Q}^{II}$ Lagrange multiplier $\lambda^{Q}$ can be evaluated which   ensures appearance of no further constraints in this theory. Now, we have found all the constraints. The constraints eq.(\ref{CE:eq}) shows that any field $A_{n}$  can be written in terms of the other fields, 
\begin{eqnarray}
A_{n}=A_{n}(Q,A_{j},C-B_{1}).
\end{eqnarray}
After taking time derivatives of $A_{n},$ we get
\begin{eqnarray}
\dot{A_{n}}=\frac{\partial A_{n}}{\partial Q} \dot{Q}+\frac{\partial A_{n}}{\partial A_{j}} \dot{A_{j}}+\frac{\partial A_{n}}{\partial C} \dot{C}-\frac{\partial A_{n}}{\partial B_{1}}\dot{B_{1}}\label{eq:yt}
\end{eqnarray}
Replacing the time derivative terms of eq.(\ref{eq:yt})  by their corresponding momenta and ignoring the spatial derivative terms, then we obtain,
\begin{eqnarray}
P_{B_{n}}=\frac{\partial A_{n}}{\partial Q} \dot{Q}+\frac{\partial A_{n}}{\partial A_{j}} P_{B_{j}}+\frac{\partial A_{n}}{\partial C} \frac{1}{3}(\Pi-C P_{C})-\frac{\partial A_{n}}{\partial B_{1}} P_{A_{1}}\label{SE:eq}
\end{eqnarray}
Whereas the time derivative of Q can be fixed by the equation $\dot{Q}=\frac{\partial H}{\partial P_{Q}}=\lambda_{Q}.$ Then eq.(\ref{SE:eq}),
yields,
\begin{eqnarray}
P_{B_{n}}=\frac{\partial A_{n}}{\partial Q} \lambda_{Q}+\frac{\partial A_{n}}{\partial A_{j}} P_{B_{j}}+\frac{\partial A_{n}}{\partial C} \frac{1}{3}(\Pi-C P_{C})-\frac{\partial A_{n}}{\partial B_{1}} P_{A_{1}}\label{SE:eq1}
\end{eqnarray}
Finally we express $P_{B_{n}}$ in terms of other variables. After putting $P_{B_{n}}$ in eq.(\ref{CE2:eq}), we get
\begin{eqnarray}
\Xi_{Q}^{II} = N \Bigg[\frac{1}{3}(\Pi-C P_{C})\left(1+ \frac{\partial f'}{\partial A_{n}}\frac{\partial A_{n}}{\partial C}\right) +P_{A_{1}}\left(1- \frac{\partial f'}{\partial A_{n}}\frac{\partial A_{n}}{\partial B_{1}}\right) \nonumber\\-\left(\frac{\partial f'}{\partial A_{1}}+ \frac{\partial f'}{\partial A_{n}}\frac{\partial A_{n}}{\partial A_{1}}\right)P_{B_{1}}+....+\left(\frac{\partial f'}{\partial A_{j}}+ \frac{\partial f'}{\partial A_{n}}\frac{\partial A_{n}}{\partial A_{j}}\right)P_{B_{j}} \\\nonumber+.... \left(\frac{\partial f'}{\partial A_{n-1}}+ \frac{\partial f'}{\partial A_{n}}\frac{\partial A_{n}}{\partial A_{n-1}}\right)P_{B_{n-1}}\Bigg]  + \sqrt{h}\lambda^{Q}\left( f''+N\frac{\partial A_{n}}{\partial Q} \frac{\partial f'}{\partial A_{n}}\right)\approx0.
\end{eqnarray}
This constraint $\Xi_{Q}^{II}\approx 0$ has a solution of the form
\begin{eqnarray}
\frac{\partial A_{n}}{\partial C}=& -\frac{1}{ \frac{\partial f'}{\partial A_{n}}}=-\frac{1}{f_{,QA_{n}}}\label{eq:75}
\end{eqnarray}
\begin{eqnarray}
 \frac{\partial A_{n}}{\partial A_{j}}=&-\frac{\frac{\partial f'}{\partial A_{j}}}{\frac{\partial f'(Q)}{\partial A_{n}}}=-\frac{f_{,QA_{j}}}{f_{,QA_{n}}}\\
 \frac{\partial A_{n}}{\partial B_{1}}=&-\frac{1}{ \frac{\partial f'}{\partial A_{n}}}=\frac{1}{f_{,QA_{n}}}\\
N\frac{\partial A_{n}}{\partial Q} =&\ \ \frac{f''}{\frac{\partial f'}{\partial A_{n}}}=\frac{f_{,QQ}}{f_{,QA_{n}}}\label{eq:as}
\end{eqnarray}
Taking the result from eq.(\ref{eq:75}-\ref{eq:as}) then the eq.(\ref{SE:eq1})  becomes,
\begin{eqnarray}
P_{B_{n}}=\frac{f_{,QQ}}{f_{,QA_{n}}} \lambda_{Q}+-\frac{f_{,QA_{j}}}{f_{,QA_{n}}} P_{B_{j}}-\frac{1}{f_{,QA_{n}}} \frac{1}{3}(\Pi-C P_{C})-\frac{1}{f_{,QA_{n}}} P_{A_{1}}
\end{eqnarray}
Replacing $P_{B_{n}}$ by the above expression  in eq.(\ref{HN:eq}), we obtain
\begin{eqnarray}
\begin{aligned}
\mathcal{H}_{N} & = \frac{1}{\sqrt{h} C} \Pi^{ij} h_{ik} h_{jl} \Pi^{kl} -\frac{1}{3\sqrt{h} C} \Pi^{2} -\frac{1}{3\sqrt{h}} \Pi P_{C} + \frac{1}{6 \sqrt{h}} C P_{C}^{2} -\sqrt{h} C \mathcal{R} \\& - \sqrt{h}  f(Q,A_{1},A_{2},.......A_{n})-Q(B_{1}-C)-\mathcal{D}_{i}\mathcal{D}^{i}C\\&  +  \sum_{k= 1}^{n-1}  B_{k} A_{k+1}  - h^{ij} \sum_{n=1}^{k}  \mathcal{D}_{i}B_{n}\mathcal{D}_{j}A_{n} + \sum_{k=1}^{n-1}  P_{B_{n}}P_{A_{n}}\\& +P_{A_{n}}\left(\frac{f_{,QQ}}{f_{,QA_{n}}} \lambda_{Q}+-\frac{f_{,QA_{j}}}{f_{,QA_{n}}} P_{B_{j}}-\frac{1}{f_{,QA_{n}}} \frac{1}{3}(\Pi-C P_{C})-\frac{1}{f_{,QA_{n}}} P_{A_{1}}\right),
\end{aligned}
\end{eqnarray}

Finally, we arrive at the final form of Hamiltonian after exhausting all the constraints. Various linear momenta terms still exist in this Hamiltonian, which indicates the presence of the Ostrogradsky ghost in this theory.

\subsection{Hamiltonian Formalism for \texorpdfstring{$Rf(R,\Box^{-1}R,\Box^{-2}R...\Box^{-n}R)$}{TEXT}}
In this section, we follow the analysis of previous section for deriving Hamiltonian of the Lagrangian in the action,
\begin{eqnarray}
S=\int d^{3}x \sqrt{-g}\ R f(\Box^{-1}R,\Box^{-2}R..........\Box^{-n}R), \ \ \ \textnormal{with} \ \ \  \ n \neq +\infty, \label{Ac:eq3}
\end{eqnarray}
and its scalar tensor form looks,
\begin{eqnarray}
S=\int d^{3}x\sqrt{-g}\bigg[Q f(A_{1},A_{2},.......A_{n})+C(R-Q)+B_{1}(Q-\Box A_{1})\\\nonumber+B_{2}(A_{1}-\Box A_{2}),.........B_{n}(A_{n-1}-\Box A_{n})\bigg].
\end{eqnarray}
Action eq.(\ref{Ac:eq3}) in a expanded form is,
\begin{eqnarray}
S=\int d^{3}x \sqrt{-g}\bigg[Q f(A_{1},A_{2},.......A_{n})+CR+Q( B_{1}-C)+g^{\mu\nu}(\partial_{\mu}B_{1}\partial_{\nu} A_{1}+\\\partial_{\mu}B_{2}\partial_{\nu}A_{2}......... \partial_{\mu}B_{n}\partial_{\nu}A_{n})\nonumber+B_{2}A_{1} +B_{3}A_{2}........B_{n}A_{n-1}\bigg].
\end{eqnarray}
The above action in 3+1 decomposition can be written as,
\begin{eqnarray}
\begin{aligned}
S^{eqv}= \int d^{3}x N\sqrt{h} \bigg [Q f(A_{1},A_{2},.......A_{n})  + C\left( K_{ij}K^{ij} -K^{2} + \mathcal{R} -Q\right) &\\ -2K \ (n^{a} \nabla_{a}C)+Q(B_{1}-C)+\mathcal{D}_{i}\mathcal{D}^{i}C  -  \sum_{k= 1}^{n-1}  B_{k} A_{k+1} &\\ + h^{ij} \sum_{n=1}^{k}  \mathcal{D}_{i}B_{n}\mathcal{D}_{j}A_{n} - \sum_{n=1}^{k}  (n^{a}\nabla_{a}B_{n})(n^{b}\nabla_{b}A_{n}) \bigg].
\end{aligned}
\end{eqnarray}

 Here canonical momenta with respect to set  $(N, N^i, h_{ij}, Q, C, A_{L}, B_{L})$ are, 
\begin{eqnarray}
\begin{aligned}
\Pi_{N} & \approx 0, \ \ \ \ \ 
\Pi_{i} \approx 0,\ \ \ 
\Pi^{ij} = \sqrt{h} C( \mathcal{K}^{ij} -h^{ij}\mathcal{K} ) - \sqrt{h} h^{ij} (n^{a} \nabla_{a}C), \qquad \qquad \qquad \qquad   \\
P_{Q} & \approx  0 , \qquad \qquad
P_{\Phi} = -2\sqrt{h} \mathcal{K}, \qquad\\
P_{A_{1}} &= \sqrt{h} (n^{a}\nabla_{a}B_{1}), \ \ P_{A_{2}} = (n^{a}\nabla_{a}B_{2}) \  ,,,,,,,\ P_{A_{k}}=\sqrt{h} (n^{a}\nabla_{a}B_{k})
\\ P_{B_{1}}& = \sqrt{h}  (n^{a}\nabla_{a}A_{1})\  \ P_{B_{2}} = \sqrt{h}  (n^{a}\nabla_{a}A_{2}),,,,,,,,P_{B_{k}} = \sqrt{h}  (n^{a}\nabla_{a}A_{k}).
\end{aligned}
\end{eqnarray}
After some simplification Hamiltonian takes a form,
\begin{equation}
\mathcal{H}= N\mathcal{H}_{N}  + N^{i} \mathcal{H}_{i,} \label{eq:dwa}
\end{equation}
with
\begin{eqnarray}
\begin{aligned}
\mathcal{H}_{N} & = \frac{1}{\sqrt{h} C} \Pi^{ij} h_{ik} h_{jl} \Pi^{kl} -\frac{1}{3\sqrt{h} C} \Pi^{2} -\frac{1}{3\sqrt{h}} \Pi P_{C} + \frac{1}{6 \sqrt{h}} C P_{C}^{2} -\sqrt{h} C \mathcal{R} \\& - \sqrt{h}   Qf(A_{1},A_{2},.......A_{n})-Q(B_{1}-C)-\mathcal{D}_{i}\mathcal{D}^{i}C\\&  +  \sum_{k= 1}^{n-1}  B_{k} A_{k+1}  - h^{ij} \sum_{n=1}^{k}  \mathcal{D}_{i}B_{n}\mathcal{D}_{j}A_{n} + \sum_{n=1}^{k}  P_{B_{n}}P_{A_{n}},
\end{aligned}
\end{eqnarray}
and
\begin{equation}
\begin{split}
\mathcal{H}_{i}  = -2 h_{ik}\mathcal{D}_{j}\Pi^{kj} + \sum_{n = 1}^{k} \Big(P_{A_{n}}\mathcal{D}_{i}A_{n} + P_{B_{n}}\mathcal{D}_{i}B_{n} \Big) + P_{\Phi}\mathcal{D}_{i}\Phi.
\end{split}
\end{equation} 
Here the constraint is 
\begin{eqnarray}
\Xi_{Q} = \partial_{t}P_{Q} =\{ P_{Q}, \mathcal{H}_{tot}\} = N\sqrt{h}\Big\{B_{1}-C -f(A_{1},A_{2}......A_{n}\Big\}\approx 0. \label{CE:eq1}
\end{eqnarray}  
and its evolution yields,
\begin{eqnarray}
\Xi_{Q}^{II}=\dot{\Xi}_{Q} = N \Bigg[\frac{1}{3}(\Pi-C P_{C}) +P_{A_{1}} -\frac{\partial f}{\partial A_{1}}P_{B_{1}}+\frac{\partial f}{\partial A_{2}}P_{B_{2}}+....\frac{\partial f}{\partial A_{j}}P_{B_{j}} \\\nonumber+.... \frac{\partial f}{\partial A_{n}}P_{B_{n}}\Bigg]\approx0.\label{CE:eq2}
\end{eqnarray} 

Equivalently, the eq.(\ref{CE:eq2}) can be written as,
\begin{eqnarray}
\Xi_{Q}^{2I}=N \Bigg[\frac{1}{3}(\Pi-C P_{C}) + -\frac{\partial f}{\partial A_{1}}P_{B_{1}}+\zeta\Bigg]\approx0.\label{CE:eq21}
\end{eqnarray} 
where,$$\zeta=\frac{\partial f}{\partial A_{2}}P_{B_{2}}+....\frac{\partial f}{\partial A_{j}}P_{B_{j}} \\\nonumber+.... \frac{\partial f}{\partial A_{n}}P_{B_{n}}$$
$\Xi_{Q}^{2I}\approx0$ acts as a tertiary constraint and its time evolution is,
\begin{eqnarray}
\Xi_{Q}^{2II}=N \bigg[\frac{-\Pi^{ij}}{\sqrt{h}} h_{ik} h_{jl} \Pi^{kl} +\frac{1}{3\sqrt{h} C} \Pi^{2} +\frac{1}{3\sqrt{h}} \Pi P_{C} - \frac{1}{6 \sqrt{h}} C P_{C}^{2} \nonumber \\+\sqrt{h} C \mathcal{R}  - \sqrt{h}C Q -\frac{\partial f}{\partial A_{1}} Q +P_{B_{1}}\{\frac{\partial f}{\partial A_{1}},\mathcal{H}\} +\{\zeta,\mathcal{H}\} \bigg]\approx0.\label{same2}
\end{eqnarray} 
Further the time evolution of $\Xi_{Q}^{2II}$ fixes the Lagrange multiplier $\lambda_{Q}.$  Now, all the Lagrange multipliers are fixed, ensuring no further constraint appears in this theory.
Therefore, we have four secondary constraints in our theory which are,
\begin{eqnarray}
 \Xi_{Q}\approx0,\qquad \Xi_{Q}^{II}\approx0,\qquad \Xi_{Q}^{2I}\approx0,\qquad \Xi_{Q}^{2II}\approx0. \label{sc}
\end{eqnarray}
The constraint equation \ref{CE:eq1} field $A_{n}$ is related to other $A_{n}$ fields by the relation
\begin{eqnarray}
A_{n}=A_{n}(A_{j},C-B_{1}).
\end{eqnarray}
By taking the time derivatives of $A_{n},$ we obtain
\begin{eqnarray}
\dot{A_{n}}=\frac{\partial A_{n}}{\partial A_{j}} \dot{A_{j}}+\frac{\partial A_{n}}{\partial C} \dot{C}-\frac{\partial A_{n}}{\partial B_{1}}\dot{B_{1}}\label{kl}
\end{eqnarray}
Eq.(\ref{kl})  can be written in terms of momenta,
\begin{eqnarray}
P_{B_{n}}=\frac{\partial A_{n}}{\partial A_{j}} P_{B_{j}}+\frac{\partial A_{n}}{\partial C} \frac{1}{3}(\Pi-C P_{C})-\frac{\partial A_{n}}{\partial B_{1}} P_{A_{1}}. \label{eq:ss}
\end{eqnarray}
 Replacing $P_{B_{n}}$ in eq.(\ref{CE:eq2}) by using eq.(\ref{eq:ss}), we get
\begin{eqnarray}
\Xi_{Q}^{I} = N \Bigg[\frac{1}{3}(\Pi-C P_{C})\left(1+ \frac{\partial f}{\partial A_{n}}\frac{\partial A_{n}}{\partial C}\right) +P_{A_{1}}\left(1- \frac{\partial f}{\partial A_{n}}\frac{\partial A_{n}}{\partial B_{1}}\right) \nonumber\\-\left(\frac{\partial f}{\partial A_{1}}+ \frac{\partial f}{\partial A_{n}}\frac{\partial A_{n}}{\partial A_{1}}\right) \label{eq:swa}P_{B_{1}}+....+\left(\frac{\partial f}{\partial A_{j}}+ \frac{\partial f}{\partial A_{n}}\frac{\partial A_{n}}{\partial A_{j}}\right)P_{B_{j}} \\\nonumber+.... \left(\frac{\partial f}{\partial A_{n-1}}+ \frac{\partial f}{\partial A_{n}}\frac{\partial A_{n}}{\partial A_{n-1}}\right)P_{B_{n-1}}\Bigg] \approx0.
\end{eqnarray}
This constraint has a solution of the form,
\begin{eqnarray}
\frac{\partial A_{n}}{\partial C}=& -\frac{1}{ \frac{\partial f(Q)}{\partial A_{n}}}=-\frac{1}{f_{,A_{n}},}\label{eq:80}\\
 \frac{\partial A_{n}}{\partial A_{j}}=&-\frac{\frac{\partial f(Q)}{\partial A_{j}}}{\frac{\partial f'(Q)}{\partial A_{n}}}=-\frac{f_{,A_{j}}}{f_{,A_{n}}},\\
 \frac{\partial A_{n}}{\partial B_{1}}=&-\frac{1}{ \frac{\partial f(Q)}{\partial A_{n}}}=\frac{1}{f_{,A_{n}}}.\label{eq:81}
\end{eqnarray}
Taking the result from eq.(\ref{eq:80}-\ref{eq:81}) then the eq.(\ref{eq:swa}) becomes,
\begin{eqnarray}
P_{B_{n}}=-\frac{f_{,QA_{j}}}{f_{,QA_{n}}} P_{B_{j}}-\frac{1}{f_{,QA_{n}}} \frac{1}{3}(\Pi-C P_{C})+-\frac{1}{f_{,QA_{n}}} P_{A_{1}}. \label{eq:awa}
\end{eqnarray}
After putting eq.(\ref{eq:awa}) into Hamiltonian eq.(\ref{eq:dwa}) takes the form,
\begin{eqnarray}
\begin{aligned}
\mathcal{H}_{N} & = \frac{1}{\sqrt{h} C} \Pi^{ij} h_{ik} h_{jl} \Pi^{kl} -\frac{1}{3\sqrt{h} C} \Pi^{2} -\frac{1}{3\sqrt{h}} \Pi P_{C} + \frac{1}{6 \sqrt{h}} C P_{C}^{2} -\sqrt{h} C \mathcal{R} \\& - \sqrt{h}  f(Q,A_{1},A_{2},.......A_{n})-Q(B_{1}-C)-\mathcal{D}_{i}\mathcal{D}^{i}C\\&  +  \sum_{k= 1}^{n-1}  B_{k} A_{k+1}  - h^{ij} \sum_{n=1}^{k}  \mathcal{D}_{i}B_{n}\mathcal{D}_{j}A_{n} + \sum_{k=1}^{n-1}  P_{B_{n}}P_{A_{n}}\\& +P_{A_{n}}\left(-\frac{f_{,QA_{j}}}{f_{,QA_{n}}} P_{B_{j}}-\frac{1}{f_{,QA_{n}}} \frac{1}{3}(\Pi-C P_{C})-\frac{1}{f_{,QA_{n}}} P_{A_{1}}\right),
\end{aligned}
\end{eqnarray}
The final Hamiltonian form can then be obtained after exhausting all the constraints. This Hamiltonian still contains numerous linear momenta terms, which indicate the presence of the Ostrogradsky ghost.

\section{Conclusion}

There has been investigation of the instability issue in nonlocal theory, arising due to the presence of higher derivatives in the Lagrangian. Studies related to this show that it is possible to overcome the issue of  Ostrogradsky instability in infinite derivative gravity models,\cite{Biswas:2013cha, Biswas:2011ar, Conroy:2014eja, Joshi:2019cyk}.

Further, the nonlocal F(R) gravity model is investigated in  \cite{Nojiri:2019dio} and demonstrates under what conditions the issue of  Ostrogradsky instability can be resolved simply by analyzing the kinetic matrix of the localized action. The final form of a kinetic matrix is obtained by relating the Lagrange multipliers using equations of motion. In order to obtain a consistent theory, it is necessary to examine the same formalism(obtaining no ghost conditions) from a Hamiltonian point of view. It is well known that Hamiltonian of non-degenerate higher derivative Lagrangian contains linear momenta terms, indicating the presence of Ostrogradsky ghost, and possible way to tackle ghost issue in Hamiltonian, first, to find any hidden condition(if any) that can relate both of linear momenta term, and second by adding a constraint. Here, we use the first method.

Our analysis starts with non-local action eq.(\ref{action}) of general functions of Ricci scalar. We derive the Hamiltonian and constraint for eq.(\ref{action}). By analyzing the constraint structure, we showe the theory that the ghost issue cannot be resolved with $F(R)\neq H(R)$. Next, we examine for the case when $F(R)$ and $H(R)$ are linear functions of Ricci scalar and notice that for $k\geq 2$ ghost issue is always present. However, for the case $k=1$, we show that no ghost appears.

 Further we analyze the case when $F(R)=H(R)$ and $G(R)=R$ for two choices $k=1$ and $k=2.$ We can obtain a ghost free theory for $k=1,$  whereas  ghost issue is always present for $k\geq 2.$
 However in \cite{Nojiri:2019dio} it is shown that that by modifying the kinetic term in a particular fashion such that no ghost appears in later choice. When we analyze the  Hamiltonian in this case we get the same ghost-free condition as in \cite{Nojiri:2019dio}.
 
 For completeness, we also study the Hamiltonian of most generalised non local models containing only Ricci scalars and find the structure of constraints and nature of ghosts in general.

Our model can be generalized for the more complicated form of the $F(R)$ gravity model.

\bibliography{refs}

\begin{thebibliography}{48}%
\makeatletter
\providecommand \@ifxundefined [1]{%
 \@ifx{#1\undefined}
}%
\providecommand \@ifnum [1]{%
 \ifnum #1\expandafter \@firstoftwo
 \else \expandafter \@secondoftwo
 \fi
}%
\providecommand \@ifx [1]{%
 \ifx #1\expandafter \@firstoftwo
 \else \expandafter \@secondoftwo
 \fi
}%
\providecommand \natexlab [1]{#1}%
\providecommand \enquote  [1]{``#1''}%
\providecommand \bibnamefont  [1]{#1}%
\providecommand \bibfnamefont [1]{#1}%
\providecommand \citenamefont [1]{#1}%
\providecommand \href@noop [0]{\@secondoftwo}%
\providecommand \href [0]{\begingroup \@sanitize@url \@href}%
\providecommand \@href[1]{\@@startlink{#1}\@@href}%
\providecommand \@@href[1]{\endgroup#1\@@endlink}%
\providecommand \@sanitize@url [0]{\catcode `\\12\catcode `\$12\catcode
  `\&12\catcode `\#12\catcode `\^12\catcode `\_12\catcode `\%12\relax}%
\providecommand \@@startlink[1]{}%
\providecommand \@@endlink[0]{}%
\providecommand \url  [0]{\begingroup\@sanitize@url \@url }%
\providecommand \@url [1]{\endgroup\@href {#1}{\urlprefix }}%
\providecommand \urlprefix  [0]{URL }%
\providecommand \Eprint [0]{\href }%
\providecommand \doibase [0]{http://dx.doi.org/}%
\providecommand \selectlanguage [0]{\@gobble}%
\providecommand \bibinfo  [0]{\@secondoftwo}%
\providecommand \bibfield  [0]{\@secondoftwo}%
\providecommand \translation [1]{[#1]}%
\providecommand \BibitemOpen [0]{}%
\providecommand \bibitemStop [0]{}%
\providecommand \bibitemNoStop [0]{.\EOS\space}%
\providecommand \EOS [0]{\spacefactor3000\relax}%
\providecommand \BibitemShut  [1]{\csname bibitem#1\endcsname}%
\let\auto@bib@innerbib\@empty
\bibitem [{\citenamefont {Riess}\ \emph {et~al.}(1998)\citenamefont {Riess}
  \emph {et~al.}}]{RiessAdamG:1998cb}%
  \BibitemOpen
  \bibfield  {author} {\bibinfo {author} {\bibfnamefont {A.~G.}\ \bibnamefont
  {Riess}} \emph {et~al.} (\bibinfo {collaboration} {Supernova Search Team}),\
  }\href {\doibase 10.1086/300499} {\bibfield  {journal} {\bibinfo  {journal}
  {Astron. J.}\ }\textbf {\bibinfo {volume} {116}},\ \bibinfo {pages} {1009}
  (\bibinfo {year} {1998})},\ \Eprint {http://arxiv.org/abs/astro-ph/9805201}
  {arXiv:astro-ph/9805201} \BibitemShut {NoStop}%
\bibitem [{\citenamefont {Spergel}\ \emph {et~al.}(2003)\citenamefont {Spergel}
  \emph {et~al.}}]{WMAP:2003elm}%
  \BibitemOpen
  \bibfield  {author} {\bibinfo {author} {\bibfnamefont {D.~N.}\ \bibnamefont
  {Spergel}} \emph {et~al.} (\bibinfo {collaboration} {WMAP}),\ }\href
  {\doibase 10.1086/377226} {\bibfield  {journal} {\bibinfo  {journal}
  {Astrophys. J. Suppl.}\ }\textbf {\bibinfo {volume} {148}},\ \bibinfo {pages}
  {175} (\bibinfo {year} {2003})},\ \Eprint
  {http://arxiv.org/abs/astro-ph/0302209} {arXiv:astro-ph/0302209} \BibitemShut
  {NoStop}%
\bibitem [{\citenamefont {Spergel}\ \emph {et~al.}(2007)\citenamefont {Spergel}
  \emph {et~al.}}]{WMAP:2006bqn}%
  \BibitemOpen
  \bibfield  {author} {\bibinfo {author} {\bibfnamefont {D.~N.}\ \bibnamefont
  {Spergel}} \emph {et~al.} (\bibinfo {collaboration} {WMAP}),\ }\href
  {\doibase 10.1086/513700} {\bibfield  {journal} {\bibinfo  {journal}
  {Astrophys. J. Suppl.}\ }\textbf {\bibinfo {volume} {170}},\ \bibinfo {pages}
  {377} (\bibinfo {year} {2007})},\ \Eprint
  {http://arxiv.org/abs/astro-ph/0603449} {arXiv:astro-ph/0603449} \BibitemShut
  {NoStop}%
\bibitem [{\citenamefont {Komatsu}\ \emph {et~al.}(2009)\citenamefont {Komatsu}
  \emph {et~al.}}]{WMAP:2008lyn}%
  \BibitemOpen
  \bibfield  {author} {\bibinfo {author} {\bibfnamefont {E.}~\bibnamefont
  {Komatsu}} \emph {et~al.} (\bibinfo {collaboration} {WMAP}),\ }\href
  {\doibase 10.1088/0067-0049/180/2/330} {\bibfield  {journal} {\bibinfo
  {journal} {Astrophys. J. Suppl.}\ }\textbf {\bibinfo {volume} {180}},\
  \bibinfo {pages} {330} (\bibinfo {year} {2009})},\ \Eprint
  {http://arxiv.org/abs/0803.0547} {arXiv:0803.0547 [astro-ph]} \BibitemShut
  {NoStop}%
\bibitem [{\citenamefont {Tegmark}\ \emph {et~al.}(2004)\citenamefont {Tegmark}
  \emph {et~al.}}]{SDSS:2003eyi}%
  \BibitemOpen
  \bibfield  {author} {\bibinfo {author} {\bibfnamefont {M.}~\bibnamefont
  {Tegmark}} \emph {et~al.} (\bibinfo {collaboration} {SDSS}),\ }\href
  {\doibase 10.1103/PhysRevD.69.103501} {\bibfield  {journal} {\bibinfo
  {journal} {Phys. Rev. D}\ }\textbf {\bibinfo {volume} {69}},\ \bibinfo
  {pages} {103501} (\bibinfo {year} {2004})},\ \Eprint
  {http://arxiv.org/abs/astro-ph/0310723} {arXiv:astro-ph/0310723} \BibitemShut
  {NoStop}%
\bibitem [{\citenamefont {Jain}\ and\ \citenamefont
  {Taylor}(2003)}]{Jain:2003tba}%
  \BibitemOpen
  \bibfield  {author} {\bibinfo {author} {\bibfnamefont {B.}~\bibnamefont
  {Jain}}\ and\ \bibinfo {author} {\bibfnamefont {A.}~\bibnamefont {Taylor}},\
  }\href {\doibase 10.1103/PhysRevLett.91.141302} {\bibfield  {journal}
  {\bibinfo  {journal} {Phys. Rev. Lett.}\ }\textbf {\bibinfo {volume} {91}},\
  \bibinfo {pages} {141302} (\bibinfo {year} {2003})},\ \Eprint
  {http://arxiv.org/abs/astro-ph/0306046} {arXiv:astro-ph/0306046} \BibitemShut
  {NoStop}%
\bibitem [{\citenamefont {Weinberg}(1989)}]{Weinberg:1988cp}%
  \BibitemOpen
  \bibfield  {author} {\bibinfo {author} {\bibfnamefont {S.}~\bibnamefont
  {Weinberg}},\ }\href {\doibase 10.1103/RevModPhys.61.1} {\bibfield  {journal}
  {\bibinfo  {journal} {Rev. Mod. Phys.}\ }\textbf {\bibinfo {volume} {61}},\
  \bibinfo {pages} {1} (\bibinfo {year} {1989})}\BibitemShut {NoStop}%
\bibitem [{\citenamefont {Capozziello}(2002)}]{Capozziello:2002rd}%
  \BibitemOpen
  \bibfield  {author} {\bibinfo {author} {\bibfnamefont {S.}~\bibnamefont
  {Capozziello}},\ }\href {\doibase 10.1142/S0218271802002025} {\bibfield
  {journal} {\bibinfo  {journal} {Int. J. Mod. Phys. D}\ }\textbf {\bibinfo
  {volume} {11}},\ \bibinfo {pages} {483} (\bibinfo {year} {2002})},\ \Eprint
  {http://arxiv.org/abs/gr-qc/0201033} {arXiv:gr-qc/0201033} \BibitemShut
  {NoStop}%
\bibitem [{\citenamefont {Nojiri}\ and\ \citenamefont
  {Odintsov}(2006)}]{Nojiri:2006gh}%
  \BibitemOpen
  \bibfield  {author} {\bibinfo {author} {\bibfnamefont {S.}~\bibnamefont
  {Nojiri}}\ and\ \bibinfo {author} {\bibfnamefont {S.~D.}\ \bibnamefont
  {Odintsov}},\ }\href {\doibase 10.1103/PhysRevD.74.086005} {\bibfield
  {journal} {\bibinfo  {journal} {Phys. Rev. D}\ }\textbf {\bibinfo {volume}
  {74}},\ \bibinfo {pages} {086005} (\bibinfo {year} {2006})},\ \Eprint
  {http://arxiv.org/abs/hep-th/0608008} {arXiv:hep-th/0608008} \BibitemShut
  {NoStop}%
\bibitem [{\citenamefont {Sotiriou}\ and\ \citenamefont
  {Faraoni}(2010)}]{Sotiriou:2008rp}%
  \BibitemOpen
  \bibfield  {author} {\bibinfo {author} {\bibfnamefont {T.~P.}\ \bibnamefont
  {Sotiriou}}\ and\ \bibinfo {author} {\bibfnamefont {V.}~\bibnamefont
  {Faraoni}},\ }\href {\doibase 10.1103/RevModPhys.82.451} {\bibfield
  {journal} {\bibinfo  {journal} {Rev. Mod. Phys.}\ }\textbf {\bibinfo {volume}
  {82}},\ \bibinfo {pages} {451} (\bibinfo {year} {2010})},\ \Eprint
  {http://arxiv.org/abs/0805.1726} {arXiv:0805.1726 [gr-qc]} \BibitemShut
  {NoStop}%
\bibitem [{\citenamefont {De~Felice}\ and\ \citenamefont
  {Tsujikawa}(2010)}]{DeFelice:2010aj}%
  \BibitemOpen
  \bibfield  {author} {\bibinfo {author} {\bibfnamefont {A.}~\bibnamefont
  {De~Felice}}\ and\ \bibinfo {author} {\bibfnamefont {S.}~\bibnamefont
  {Tsujikawa}},\ }\href {\doibase 10.12942/lrr-2010-3} {\bibfield  {journal}
  {\bibinfo  {journal} {Living Rev. Rel.}\ }\textbf {\bibinfo {volume} {13}},\
  \bibinfo {pages} {3} (\bibinfo {year} {2010})},\ \Eprint
  {http://arxiv.org/abs/1002.4928} {arXiv:1002.4928 [gr-qc]} \BibitemShut
  {NoStop}%
\bibitem [{\citenamefont {Starobinsky}(2007)}]{Starobinsky:2007hu}%
  \BibitemOpen
  \bibfield  {author} {\bibinfo {author} {\bibfnamefont {A.~A.}\ \bibnamefont
  {Starobinsky}},\ }\href {\doibase 10.1134/S0021364007150027} {\bibfield
  {journal} {\bibinfo  {journal} {JETP Lett.}\ }\textbf {\bibinfo {volume}
  {86}},\ \bibinfo {pages} {157} (\bibinfo {year} {2007})},\ \Eprint
  {http://arxiv.org/abs/0706.2041} {arXiv:0706.2041 [astro-ph]} \BibitemShut
  {NoStop}%
\bibitem [{\citenamefont {Odintsov}\ and\ \citenamefont
  {Oikonomou}(2019)}]{Odintsov:2019evb}%
  \BibitemOpen
  \bibfield  {author} {\bibinfo {author} {\bibfnamefont {S.~D.}\ \bibnamefont
  {Odintsov}}\ and\ \bibinfo {author} {\bibfnamefont {V.~K.}\ \bibnamefont
  {Oikonomou}},\ }\href {\doibase 10.1103/PhysRevD.99.104070} {\bibfield
  {journal} {\bibinfo  {journal} {Phys. Rev. D}\ }\textbf {\bibinfo {volume}
  {99}},\ \bibinfo {pages} {104070} (\bibinfo {year} {2019})},\ \Eprint
  {http://arxiv.org/abs/1905.03496} {arXiv:1905.03496 [gr-qc]} \BibitemShut
  {NoStop}%
\bibitem [{\citenamefont {Dirian}\ \emph {et~al.}(2014)\citenamefont {Dirian},
  \citenamefont {Foffa}, \citenamefont {Khosravi}, \citenamefont {Kunz},\ and\
  \citenamefont {Maggiore}}]{Dirian:2014ara}%
  \BibitemOpen
  \bibfield  {author} {\bibinfo {author} {\bibfnamefont {Y.}~\bibnamefont
  {Dirian}}, \bibinfo {author} {\bibfnamefont {S.}~\bibnamefont {Foffa}},
  \bibinfo {author} {\bibfnamefont {N.}~\bibnamefont {Khosravi}}, \bibinfo
  {author} {\bibfnamefont {M.}~\bibnamefont {Kunz}}, \ and\ \bibinfo {author}
  {\bibfnamefont {M.}~\bibnamefont {Maggiore}},\ }\href {\doibase
  10.1088/1475-7516/2014/06/033} {\bibfield  {journal} {\bibinfo  {journal}
  {JCAP}\ }\textbf {\bibinfo {volume} {06}},\ \bibinfo {pages} {033} (\bibinfo
  {year} {2014})},\ \Eprint {http://arxiv.org/abs/1403.6068} {arXiv:1403.6068
  [astro-ph.CO]} \BibitemShut {NoStop}%
\bibitem [{\citenamefont {Belgacem}\ \emph {et~al.}(2020)\citenamefont
  {Belgacem}, \citenamefont {Dirian}, \citenamefont {Finke}, \citenamefont
  {Foffa},\ and\ \citenamefont {Maggiore}}]{Belgacem:2020pdz}%
  \BibitemOpen
  \bibfield  {author} {\bibinfo {author} {\bibfnamefont {E.}~\bibnamefont
  {Belgacem}}, \bibinfo {author} {\bibfnamefont {Y.}~\bibnamefont {Dirian}},
  \bibinfo {author} {\bibfnamefont {A.}~\bibnamefont {Finke}}, \bibinfo
  {author} {\bibfnamefont {S.}~\bibnamefont {Foffa}}, \ and\ \bibinfo {author}
  {\bibfnamefont {M.}~\bibnamefont {Maggiore}},\ }\href {\doibase
  10.1088/1475-7516/2020/04/010} {\bibfield  {journal} {\bibinfo  {journal}
  {JCAP}\ }\textbf {\bibinfo {volume} {04}},\ \bibinfo {pages} {010} (\bibinfo
  {year} {2020})},\ \Eprint {http://arxiv.org/abs/2001.07619} {arXiv:2001.07619
  [astro-ph.CO]} \BibitemShut {NoStop}%
\bibitem [{\citenamefont {Wetterich}(1998)}]{Wetterich:1997bz}%
  \BibitemOpen
  \bibfield  {author} {\bibinfo {author} {\bibfnamefont {C.}~\bibnamefont
  {Wetterich}},\ }\href {\doibase 10.1023/A:1018837319976} {\bibfield
  {journal} {\bibinfo  {journal} {Gen. Rel. Grav.}\ }\textbf {\bibinfo {volume}
  {30}},\ \bibinfo {pages} {159} (\bibinfo {year} {1998})},\ \Eprint
  {http://arxiv.org/abs/gr-qc/9704052} {arXiv:gr-qc/9704052} \BibitemShut
  {NoStop}%
\bibitem [{\citenamefont {Deser}\ and\ \citenamefont
  {Woodard}(2007)}]{Deser:2007jk}%
  \BibitemOpen
  \bibfield  {author} {\bibinfo {author} {\bibfnamefont {S.}~\bibnamefont
  {Deser}}\ and\ \bibinfo {author} {\bibfnamefont {R.~P.}\ \bibnamefont
  {Woodard}},\ }\href {\doibase 10.1103/PhysRevLett.99.111301} {\bibfield
  {journal} {\bibinfo  {journal} {Phys. Rev. Lett.}\ }\textbf {\bibinfo
  {volume} {99}},\ \bibinfo {pages} {111301} (\bibinfo {year} {2007})},\
  \Eprint {http://arxiv.org/abs/0706.2151} {arXiv:0706.2151 [astro-ph]}
  \BibitemShut {NoStop}%
\bibitem [{\citenamefont {Capozziello}\ and\ \citenamefont
  {Bajardi}(2022)}]{Capozziello:2022lic}%
  \BibitemOpen
  \bibfield  {author} {\bibinfo {author} {\bibfnamefont {S.}~\bibnamefont
  {Capozziello}}\ and\ \bibinfo {author} {\bibfnamefont {F.}~\bibnamefont
  {Bajardi}},\ }\href {\doibase 10.1142/S0218271822300099} {\bibfield
  {journal} {\bibinfo  {journal} {International Journal of Modern Physics D}\ }
  (\bibinfo {year} {2022}),\ 10.1142/S0218271822300099},\ \Eprint
  {http://arxiv.org/abs/2201.04512} {arXiv:2201.04512 [gr-qc]} \BibitemShut
  {NoStop}%
\bibitem [{\citenamefont {M.Ostrogradsky}(1850)}]{Ostrogradsky:1850fid}%
  \BibitemOpen
  \bibfield  {author} {\bibinfo {author} {\bibnamefont {M.Ostrogradsky}},\
  }\href@noop {} {\bibfield  {journal} {\bibinfo  {journal}
  {Mem.Acad.St.Petersbourg}\ }\textbf {\bibinfo {volume} {6}} (\bibinfo {year}
  {1850})}\BibitemShut {NoStop}%
\bibitem [{\citenamefont {Woodard}(2015)}]{Woodard:2015zca}%
  \BibitemOpen
  \bibfield  {author} {\bibinfo {author} {\bibfnamefont {R.~P.}\ \bibnamefont
  {Woodard}},\ }\href {\doibase 10.4249/scholarpedia.32243} {\bibfield
  {journal} {\bibinfo  {journal} {Scholarpedia}\ }\textbf {\bibinfo {volume}
  {10}},\ \bibinfo {pages} {32243} (\bibinfo {year} {2015})},\ \Eprint
  {http://arxiv.org/abs/1506.02210} {arXiv:1506.02210 [hep-th]} \BibitemShut
  {NoStop}%
\bibitem [{\citenamefont {Horndeski}(1974)}]{Horndeski:1974wa}%
  \BibitemOpen
  \bibfield  {author} {\bibinfo {author} {\bibfnamefont {G.~W.}\ \bibnamefont
  {Horndeski}},\ }\href {\doibase 10.1007/BF01807638} {\bibfield  {journal}
  {\bibinfo  {journal} {Int. J. Theor. Phys.}\ }\textbf {\bibinfo {volume}
  {10}},\ \bibinfo {pages} {363} (\bibinfo {year} {1974})}\BibitemShut
  {NoStop}%
\bibitem [{\citenamefont {Nicolis}\ \emph {et~al.}(2009)\citenamefont
  {Nicolis}, \citenamefont {Rattazzi},\ and\ \citenamefont
  {Trincherini}}]{Nicolis:2008in}%
  \BibitemOpen
  \bibfield  {author} {\bibinfo {author} {\bibfnamefont {A.}~\bibnamefont
  {Nicolis}}, \bibinfo {author} {\bibfnamefont {R.}~\bibnamefont {Rattazzi}}, \
  and\ \bibinfo {author} {\bibfnamefont {E.}~\bibnamefont {Trincherini}},\
  }\href {\doibase 10.1103/PhysRevD.79.064036} {\bibfield  {journal} {\bibinfo
  {journal} {Phys. Rev. D}\ }\textbf {\bibinfo {volume} {79}},\ \bibinfo
  {pages} {064036} (\bibinfo {year} {2009})},\ \Eprint
  {http://arxiv.org/abs/0811.2197} {arXiv:0811.2197 [hep-th]} \BibitemShut
  {NoStop}%
\bibitem [{\citenamefont {Deffayet}\ \emph {et~al.}(2011)\citenamefont
  {Deffayet}, \citenamefont {Gao}, \citenamefont {Steer},\ and\ \citenamefont
  {Zahariade}}]{Deffayet:2011gz}%
  \BibitemOpen
  \bibfield  {author} {\bibinfo {author} {\bibfnamefont {C.}~\bibnamefont
  {Deffayet}}, \bibinfo {author} {\bibfnamefont {X.}~\bibnamefont {Gao}},
  \bibinfo {author} {\bibfnamefont {D.~A.}\ \bibnamefont {Steer}}, \ and\
  \bibinfo {author} {\bibfnamefont {G.}~\bibnamefont {Zahariade}},\ }\href
  {\doibase 10.1103/PhysRevD.84.064039} {\bibfield  {journal} {\bibinfo
  {journal} {Phys. Rev. D}\ }\textbf {\bibinfo {volume} {84}},\ \bibinfo
  {pages} {064039} (\bibinfo {year} {2011})},\ \Eprint
  {http://arxiv.org/abs/1103.3260} {arXiv:1103.3260 [hep-th]} \BibitemShut
  {NoStop}%
\bibitem [{\citenamefont {Kobayashi}\ \emph {et~al.}(2011)\citenamefont
  {Kobayashi}, \citenamefont {Yamaguchi},\ and\ \citenamefont
  {Yokoyama}}]{Kobayashi:2011nu}%
  \BibitemOpen
  \bibfield  {author} {\bibinfo {author} {\bibfnamefont {T.}~\bibnamefont
  {Kobayashi}}, \bibinfo {author} {\bibfnamefont {M.}~\bibnamefont
  {Yamaguchi}}, \ and\ \bibinfo {author} {\bibfnamefont {J.}~\bibnamefont
  {Yokoyama}},\ }\href {\doibase 10.1143/PTP.126.511} {\bibfield  {journal}
  {\bibinfo  {journal} {Prog. Theor. Phys.}\ }\textbf {\bibinfo {volume}
  {126}},\ \bibinfo {pages} {511} (\bibinfo {year} {2011})},\ \Eprint
  {http://arxiv.org/abs/1105.5723} {arXiv:1105.5723 [hep-th]} \BibitemShut
  {NoStop}%
\bibitem [{\citenamefont {Gleyzes}\ \emph
  {et~al.}(2015{\natexlab{a}})\citenamefont {Gleyzes}, \citenamefont
  {Langlois}, \citenamefont {Piazza},\ and\ \citenamefont
  {Vernizzi}}]{Gleyzes:2014dya}%
  \BibitemOpen
  \bibfield  {author} {\bibinfo {author} {\bibfnamefont {J.}~\bibnamefont
  {Gleyzes}}, \bibinfo {author} {\bibfnamefont {D.}~\bibnamefont {Langlois}},
  \bibinfo {author} {\bibfnamefont {F.}~\bibnamefont {Piazza}}, \ and\ \bibinfo
  {author} {\bibfnamefont {F.}~\bibnamefont {Vernizzi}},\ }\href {\doibase
  10.1103/PhysRevLett.114.211101} {\bibfield  {journal} {\bibinfo  {journal}
  {Phys. Rev. Lett.}\ }\textbf {\bibinfo {volume} {114}},\ \bibinfo {pages}
  {211101} (\bibinfo {year} {2015}{\natexlab{a}})},\ \Eprint
  {http://arxiv.org/abs/1404.6495} {arXiv:1404.6495 [hep-th]} \BibitemShut
  {NoStop}%
\bibitem [{\citenamefont {Gleyzes}\ \emph
  {et~al.}(2015{\natexlab{b}})\citenamefont {Gleyzes}, \citenamefont
  {Langlois}, \citenamefont {Piazza},\ and\ \citenamefont
  {Vernizzi}}]{Gleyzes:2014qga}%
  \BibitemOpen
  \bibfield  {author} {\bibinfo {author} {\bibfnamefont {J.}~\bibnamefont
  {Gleyzes}}, \bibinfo {author} {\bibfnamefont {D.}~\bibnamefont {Langlois}},
  \bibinfo {author} {\bibfnamefont {F.}~\bibnamefont {Piazza}}, \ and\ \bibinfo
  {author} {\bibfnamefont {F.}~\bibnamefont {Vernizzi}},\ }\href {\doibase
  10.1088/1475-7516/2015/02/018} {\bibfield  {journal} {\bibinfo  {journal}
  {JCAP}\ }\textbf {\bibinfo {volume} {02}},\ \bibinfo {pages} {018} (\bibinfo
  {year} {2015}{\natexlab{b}})},\ \Eprint {http://arxiv.org/abs/1408.1952}
  {arXiv:1408.1952 [astro-ph.CO]} \BibitemShut {NoStop}%
\bibitem [{\citenamefont {Langlois}\ and\ \citenamefont
  {Noui}(2016)}]{Langlois:2015cwa}%
  \BibitemOpen
  \bibfield  {author} {\bibinfo {author} {\bibfnamefont {D.}~\bibnamefont
  {Langlois}}\ and\ \bibinfo {author} {\bibfnamefont {K.}~\bibnamefont
  {Noui}},\ }\href {\doibase 10.1088/1475-7516/2016/02/034} {\bibfield
  {journal} {\bibinfo  {journal} {JCAP}\ }\textbf {\bibinfo {volume} {02}},\
  \bibinfo {pages} {034} (\bibinfo {year} {2016})},\ \Eprint
  {http://arxiv.org/abs/1510.06930} {arXiv:1510.06930 [gr-qc]} \BibitemShut
  {NoStop}%
\bibitem [{\citenamefont {Motohashi}\ \emph {et~al.}(2016)\citenamefont
  {Motohashi}, \citenamefont {Noui}, \citenamefont {Suyama}, \citenamefont
  {Yamaguchi},\ and\ \citenamefont {Langlois}}]{Motohashi:2016ftl}%
  \BibitemOpen
  \bibfield  {author} {\bibinfo {author} {\bibfnamefont {H.}~\bibnamefont
  {Motohashi}}, \bibinfo {author} {\bibfnamefont {K.}~\bibnamefont {Noui}},
  \bibinfo {author} {\bibfnamefont {T.}~\bibnamefont {Suyama}}, \bibinfo
  {author} {\bibfnamefont {M.}~\bibnamefont {Yamaguchi}}, \ and\ \bibinfo
  {author} {\bibfnamefont {D.}~\bibnamefont {Langlois}},\ }\href {\doibase
  10.1088/1475-7516/2016/07/033} {\bibfield  {journal} {\bibinfo  {journal}
  {JCAP}\ }\textbf {\bibinfo {volume} {07}},\ \bibinfo {pages} {033} (\bibinfo
  {year} {2016})},\ \Eprint {http://arxiv.org/abs/1603.09355} {arXiv:1603.09355
  [hep-th]} \BibitemShut {NoStop}%
\bibitem [{\citenamefont {Joshi}\ and\ \citenamefont
  {Panda}(2021{\natexlab{a}})}]{Joshi:2019tzr}%
  \BibitemOpen
  \bibfield  {author} {\bibinfo {author} {\bibfnamefont {P.}~\bibnamefont
  {Joshi}}\ and\ \bibinfo {author} {\bibfnamefont {S.}~\bibnamefont {Panda}},\
  }\href {\doibase 10.1007/978-981-33-4408-2_128} {\bibfield  {journal}
  {\bibinfo  {journal} {Springer Proc. Phys.}\ }\textbf {\bibinfo {volume}
  {261}},\ \bibinfo {pages} {901} (\bibinfo {year} {2021}{\natexlab{a}})},\
  \Eprint {http://arxiv.org/abs/1906.02498} {arXiv:1906.02498 [gr-qc]}
  \BibitemShut {NoStop}%
\bibitem [{\citenamefont {Joshi}\ and\ \citenamefont
  {Panda}(2021{\natexlab{b}})}]{Joshi:2021azw}%
  \BibitemOpen
  \bibfield  {author} {\bibinfo {author} {\bibfnamefont {P.}~\bibnamefont
  {Joshi}}\ and\ \bibinfo {author} {\bibfnamefont {S.}~\bibnamefont {Panda}},\
  }\href@noop {} {\  (\bibinfo {year} {2021}{\natexlab{b}})},\ \Eprint
  {http://arxiv.org/abs/2111.11791} {arXiv:2111.11791 [hep-th]} \BibitemShut
  {NoStop}%
\bibitem [{\citenamefont {Langlois}(2019)}]{Langlois:2018dxi}%
  \BibitemOpen
  \bibfield  {author} {\bibinfo {author} {\bibfnamefont {D.}~\bibnamefont
  {Langlois}},\ }\href {\doibase 10.1142/S0218271819420069} {\bibfield
  {journal} {\bibinfo  {journal} {Int. J. Mod. Phys. D}\ }\textbf {\bibinfo
  {volume} {28}},\ \bibinfo {pages} {1942006} (\bibinfo {year} {2019})},\
  \Eprint {http://arxiv.org/abs/1811.06271} {arXiv:1811.06271 [gr-qc]}
  \BibitemShut {NoStop}%
\bibitem [{\citenamefont {Kobayashi}(2019)}]{Kobayashi:2019hrl}%
  \BibitemOpen
  \bibfield  {author} {\bibinfo {author} {\bibfnamefont {T.}~\bibnamefont
  {Kobayashi}},\ }\href {\doibase 10.1088/1361-6633/ab2429} {\bibfield
  {journal} {\bibinfo  {journal} {Rept. Prog. Phys.}\ }\textbf {\bibinfo
  {volume} {82}},\ \bibinfo {pages} {086901} (\bibinfo {year} {2019})},\
  \Eprint {http://arxiv.org/abs/1901.07183} {arXiv:1901.07183 [gr-qc]}
  \BibitemShut {NoStop}%
\bibitem [{\citenamefont {Chen}\ \emph {et~al.}(2013)\citenamefont {Chen},
  \citenamefont {Fasiello}, \citenamefont {Lim},\ and\ \citenamefont
  {Tolley}}]{Chen:2012au}%
  \BibitemOpen
  \bibfield  {author} {\bibinfo {author} {\bibfnamefont {T.-j.}\ \bibnamefont
  {Chen}}, \bibinfo {author} {\bibfnamefont {M.}~\bibnamefont {Fasiello}},
  \bibinfo {author} {\bibfnamefont {E.~A.}\ \bibnamefont {Lim}}, \ and\
  \bibinfo {author} {\bibfnamefont {A.~J.}\ \bibnamefont {Tolley}},\ }\href
  {\doibase 10.1088/1475-7516/2013/02/042} {\bibfield  {journal} {\bibinfo
  {journal} {JCAP}\ }\textbf {\bibinfo {volume} {02}},\ \bibinfo {pages} {042}
  (\bibinfo {year} {2013})},\ \Eprint {http://arxiv.org/abs/1209.0583}
  {arXiv:1209.0583 [hep-th]} \BibitemShut {NoStop}%
\bibitem [{\citenamefont {Wipf}(1994)}]{Wipf:1993xg}%
  \BibitemOpen
  \bibfield  {author} {\bibinfo {author} {\bibfnamefont {A.~W.}\ \bibnamefont
  {Wipf}},\ }\href {\doibase 10.1007/3-540-58339-4_14} {\bibfield  {journal}
  {\bibinfo  {journal} {Lect. Notes Phys.}\ }\textbf {\bibinfo {volume}
  {434}},\ \bibinfo {pages} {22} (\bibinfo {year} {1994})},\ \Eprint
  {http://arxiv.org/abs/hep-th/9312078} {arXiv:hep-th/9312078} \BibitemShut
  {NoStop}%
\bibitem [{\citenamefont {Dirac}(1958{\natexlab{a}})}]{Dirac:1958sq}%
  \BibitemOpen
  \bibfield  {author} {\bibinfo {author} {\bibfnamefont {P.~A.~M.}\
  \bibnamefont {Dirac}},\ }\href {\doibase 10.1098/rspa.1958.0141} {\bibfield
  {journal} {\bibinfo  {journal} {Proc. Roy. Soc. Lond. A}\ }\textbf {\bibinfo
  {volume} {246}},\ \bibinfo {pages} {326} (\bibinfo {year}
  {1958}{\natexlab{a}})}\BibitemShut {NoStop}%
\bibitem [{\citenamefont {Dirac}(1958{\natexlab{b}})}]{Dirac:1958sc}%
  \BibitemOpen
  \bibfield  {author} {\bibinfo {author} {\bibfnamefont {P.~A.~M.}\
  \bibnamefont {Dirac}},\ }\href {\doibase 10.1098/rspa.1958.0142} {\bibfield
  {journal} {\bibinfo  {journal} {Proc. Roy. Soc. Lond. A}\ }\textbf {\bibinfo
  {volume} {246}},\ \bibinfo {pages} {333} (\bibinfo {year}
  {1958}{\natexlab{b}})}\BibitemShut {NoStop}%
\bibitem [{\citenamefont {Dirac}(2013)}]{dirac2013lectures}%
  \BibitemOpen
  \bibfield  {author} {\bibinfo {author} {\bibfnamefont {P.}~\bibnamefont
  {Dirac}},\ }\href {https://books.google.co.in/books?id=Z3XCAgAAQBAJ} {\emph
  {\bibinfo {title} {Lectures on Quantum Mechanics}}},\ Dover Books on Physics\
  (\bibinfo  {publisher} {Dover Publications},\ \bibinfo {year}
  {2013})\BibitemShut {NoStop}%
\bibitem [{\citenamefont {Matschull}(1996)}]{Matschull:1996up}%
  \BibitemOpen
  \bibfield  {author} {\bibinfo {author} {\bibfnamefont {H.-J.}\ \bibnamefont
  {Matschull}},\ }\href@noop {} {\  (\bibinfo {year} {1996})},\ \Eprint
  {http://arxiv.org/abs/quant-ph/9606031} {arXiv:quant-ph/9606031} \BibitemShut
  {NoStop}%
\bibitem [{\citenamefont {Kluson}(2011)}]{Kluson:2011tb}%
  \BibitemOpen
  \bibfield  {author} {\bibinfo {author} {\bibfnamefont {J.}~\bibnamefont
  {Kluson}},\ }\href {\doibase 10.1007/JHEP09(2011)001} {\bibfield  {journal}
  {\bibinfo  {journal} {JHEP}\ }\textbf {\bibinfo {volume} {09}},\ \bibinfo
  {pages} {001} (\bibinfo {year} {2011})},\ \Eprint
  {http://arxiv.org/abs/1105.6056} {arXiv:1105.6056 [hep-th]} \BibitemShut
  {NoStop}%
\bibitem [{\citenamefont {Joshi}\ \emph {et~al.}(2019)\citenamefont {Joshi},
  \citenamefont {Kumar},\ and\ \citenamefont {Panda}}]{Joshi:2019cyk}%
  \BibitemOpen
  \bibfield  {author} {\bibinfo {author} {\bibfnamefont {P.}~\bibnamefont
  {Joshi}}, \bibinfo {author} {\bibfnamefont {U.}~\bibnamefont {Kumar}}, \ and\
  \bibinfo {author} {\bibfnamefont {S.}~\bibnamefont {Panda}},\ }\href
  {\doibase 10.1142/S0219887822500360} {\bibfield  {journal} {\bibinfo
  {journal} {International Journal of Geometric Methods in Modern Physics}\ }
  (\bibinfo {year} {2019}),\ 10.1142/S0219887822500360},\ \Eprint
  {http://arxiv.org/abs/1909.10295} {arXiv:1909.10295 [gr-qc]} \BibitemShut
  {NoStop}%
\bibitem [{\citenamefont {De~Felice}\ and\ \citenamefont
  {Sasaki}(2015)}]{DeFelice:2014kma}%
  \BibitemOpen
  \bibfield  {author} {\bibinfo {author} {\bibfnamefont {A.}~\bibnamefont
  {De~Felice}}\ and\ \bibinfo {author} {\bibfnamefont {M.}~\bibnamefont
  {Sasaki}},\ }\href {\doibase 10.1016/j.physletb.2015.02.045} {\bibfield
  {journal} {\bibinfo  {journal} {Phys. Lett. B}\ }\textbf {\bibinfo {volume}
  {743}},\ \bibinfo {pages} {189} (\bibinfo {year} {2015})},\ \Eprint
  {http://arxiv.org/abs/1412.1575} {arXiv:1412.1575 [gr-qc]} \BibitemShut
  {NoStop}%
\bibitem [{\citenamefont {Nojiri}\ \emph {et~al.}(2020)\citenamefont {Nojiri},
  \citenamefont {Odintsov},\ and\ \citenamefont {Oikonomou}}]{Nojiri:2019dio}%
  \BibitemOpen
  \bibfield  {author} {\bibinfo {author} {\bibfnamefont {S.}~\bibnamefont
  {Nojiri}}, \bibinfo {author} {\bibfnamefont {S.~D.}\ \bibnamefont
  {Odintsov}}, \ and\ \bibinfo {author} {\bibfnamefont {V.~K.}\ \bibnamefont
  {Oikonomou}},\ }\href {\doibase 10.1016/j.dark.2020.100541} {\bibfield
  {journal} {\bibinfo  {journal} {Phys. Dark Univ.}\ }\textbf {\bibinfo
  {volume} {28}},\ \bibinfo {pages} {100541} (\bibinfo {year} {2020})},\
  \Eprint {http://arxiv.org/abs/1911.07329} {arXiv:1911.07329 [gr-qc]}
  \BibitemShut {NoStop}%
\bibitem [{\citenamefont {Arnowitt}\ \emph {et~al.}(2008)\citenamefont
  {Arnowitt}, \citenamefont {Deser},\ and\ \citenamefont
  {Misner}}]{Arnowitt:1962hi}%
  \BibitemOpen
  \bibfield  {author} {\bibinfo {author} {\bibfnamefont {R.~L.}\ \bibnamefont
  {Arnowitt}}, \bibinfo {author} {\bibfnamefont {S.}~\bibnamefont {Deser}}, \
  and\ \bibinfo {author} {\bibfnamefont {C.~W.}\ \bibnamefont {Misner}},\
  }\href {\doibase 10.1007/s10714-008-0661-1} {\bibfield  {journal} {\bibinfo
  {journal} {Gen. Rel. Grav.}\ }\textbf {\bibinfo {volume} {40}},\ \bibinfo
  {pages} {1997} (\bibinfo {year} {2008})},\ \Eprint
  {http://arxiv.org/abs/gr-qc/0405109} {arXiv:gr-qc/0405109} \BibitemShut
  {NoStop}%
\bibitem [{\citenamefont {Gourgoulhon}(2007)}]{Gourgoulhon:2007ue}%
  \BibitemOpen
  \bibfield  {author} {\bibinfo {author} {\bibfnamefont {E.}~\bibnamefont
  {Gourgoulhon}},\ }\href@noop {} {\  (\bibinfo {year} {2007})},\ \Eprint
  {http://arxiv.org/abs/gr-qc/0703035} {arXiv:gr-qc/0703035} \BibitemShut
  {NoStop}%
\bibitem [{\citenamefont {Baumgarte}\ and\ \citenamefont
  {Shapiro}(2010)}]{17b23f5522334b1ca8bb575ecaf5c01e}%
  \BibitemOpen
  \bibfield  {author} {\bibinfo {author} {\bibfnamefont {T.}~\bibnamefont
  {Baumgarte}}\ and\ \bibinfo {author} {\bibfnamefont {S.}~\bibnamefont
  {Shapiro}},\ }\href {\doibase 10.1017/CBO9781139193344} {\  (\bibinfo {year}
  {2010}),\ 10.1017/CBO9781139193344}\BibitemShut {NoStop}%
\bibitem [{\citenamefont {Biswas}\ \emph {et~al.}(2014)\citenamefont {Biswas},
  \citenamefont {Conroy}, \citenamefont {Koshelev},\ and\ \citenamefont
  {Mazumdar}}]{Biswas:2013cha}%
  \BibitemOpen
  \bibfield  {author} {\bibinfo {author} {\bibfnamefont {T.}~\bibnamefont
  {Biswas}}, \bibinfo {author} {\bibfnamefont {A.}~\bibnamefont {Conroy}},
  \bibinfo {author} {\bibfnamefont {A.~S.}\ \bibnamefont {Koshelev}}, \ and\
  \bibinfo {author} {\bibfnamefont {A.}~\bibnamefont {Mazumdar}},\ }\href
  {\doibase 10.1088/0264-9381/31/1/015022} {\bibfield  {journal} {\bibinfo
  {journal} {Class. Quant. Grav.}\ }\textbf {\bibinfo {volume} {31}},\ \bibinfo
  {pages} {015022} (\bibinfo {year} {2014})},\ \bibinfo {note} {[Erratum:
  Class.Quant.Grav. 31, 159501 (2014)]},\ \Eprint
  {http://arxiv.org/abs/1308.2319} {arXiv:1308.2319 [hep-th]} \BibitemShut
  {NoStop}%
\bibitem [{\citenamefont {Biswas}\ \emph {et~al.}(2012)\citenamefont {Biswas},
  \citenamefont {Gerwick}, \citenamefont {Koivisto},\ and\ \citenamefont
  {Mazumdar}}]{Biswas:2011ar}%
  \BibitemOpen
  \bibfield  {author} {\bibinfo {author} {\bibfnamefont {T.}~\bibnamefont
  {Biswas}}, \bibinfo {author} {\bibfnamefont {E.}~\bibnamefont {Gerwick}},
  \bibinfo {author} {\bibfnamefont {T.}~\bibnamefont {Koivisto}}, \ and\
  \bibinfo {author} {\bibfnamefont {A.}~\bibnamefont {Mazumdar}},\ }\href
  {\doibase 10.1103/PhysRevLett.108.031101} {\bibfield  {journal} {\bibinfo
  {journal} {Phys. Rev. Lett.}\ }\textbf {\bibinfo {volume} {108}},\ \bibinfo
  {pages} {031101} (\bibinfo {year} {2012})},\ \Eprint
  {http://arxiv.org/abs/1110.5249} {arXiv:1110.5249 [gr-qc]} \BibitemShut
  {NoStop}%
\bibitem [{\citenamefont {Conroy}\ \emph {et~al.}(2015)\citenamefont {Conroy},
  \citenamefont {Koivisto}, \citenamefont {Mazumdar},\ and\ \citenamefont
  {Teimouri}}]{Conroy:2014eja}%
  \BibitemOpen
  \bibfield  {author} {\bibinfo {author} {\bibfnamefont {A.}~\bibnamefont
  {Conroy}}, \bibinfo {author} {\bibfnamefont {T.}~\bibnamefont {Koivisto}},
  \bibinfo {author} {\bibfnamefont {A.}~\bibnamefont {Mazumdar}}, \ and\
  \bibinfo {author} {\bibfnamefont {A.}~\bibnamefont {Teimouri}},\ }\href
  {\doibase 10.1088/0264-9381/32/1/015024} {\bibfield  {journal} {\bibinfo
  {journal} {Class. Quant. Grav.}\ }\textbf {\bibinfo {volume} {32}},\ \bibinfo
  {pages} {015024} (\bibinfo {year} {2015})},\ \Eprint
  {http://arxiv.org/abs/1406.4998} {arXiv:1406.4998 [hep-th]} \BibitemShut
  {NoStop}%
\end{thebibliography}%

\end{document}